\documentclass[journal]{IEEEtran}
\pdfoutput=0
\usepackage{cite}
\ifCLASSINFOpdf
\else
\fi
\usepackage{bbm, graphicx, caption}
\usepackage{amsmath,graphicx,enumitem,amssymb}
\usepackage{epsfig,epic,eepic,psfrag}
\usepackage{psfrag,color,bm,graphics,subfig}
\usepackage{cite,amsthm,multirow, algorithm}
\usepackage{siunitx, booktabs}
\usepackage[table]{xcolor}
\def\x{{\mathbf x}}

\def\Ibf{{\mathbf X}}
\def\h{{h}}

\def\MS{{\mathbf{m}^{(\text{MS})}_{\h,K'}\,(\x)}}
\def\RMS{{\mathbf{m}^{(\text{R-MS})}_{\h,K'}\,(\x)}}
\def\SSMS{{\mathbf{m}^{(\text{SS-MS})}_{\h,K'}\,(\x)}}

\def\I{{\mathcal{I} }}
\newcommand{\func}[2]{#1\!\left(#2\right)\,} 

\newtheorem{mydef}{Theorem}
\begin{document}
\title{Measuring Blood Glucose Concentrations in Photometric Glucometers Requiring Very Small Sample Volumes}
\author{Nevine~Demitri,~\IEEEmembership{Student Member,~IEEE,}
Abdelhak M.~Zoubir,~\IEEEmembership{Fellow,~IEEE,}
\thanks{N.\,Demitri and A.\,M.\,Zoubir are with the Signal Processing Group, Institute of Telecommunications, Technische Universit\"at Darmstadt, Merckstr. 25, 64283 Darmstadt, Germany.}
\thanks{}}
\markboth{}%
{Shell \MakeLowercase{\textit{et al.}}: Bare Demo of IEEEtran.cls for Journals}
\maketitle
\begin{abstract}
Glucometers present an important self-monitoring tool for diabetes patients and therefore must exhibit high accuracy as well as good usability features. 
Based on an invasive, photometric measurement principle that drastically reduces the volume of the blood sample needed from the patient, we present a framework that is capable of dealing with small blood samples, while maintaining the required accuracy. 
The framework consists of two major parts: 1) image segmentation; and 2) convergence detection. 
Step 1) is based on iterative mode-seeking methods to estimate the intensity value of the region of interest. We present several variations of these methods and give theoretical proofs of their convergence. 
Our approach is able to deal with changes in the number and position of clusters without any prior knowledge. Furthermore, we propose a method based on sparse approximation to decrease the computational load, while maintaining accuracy. 
Step 2) is achieved by employing temporal tracking and prediction, herewith decreasing the measurement time, and, thus, improving usability. Our framework is tested on several real data sets with different characteristics. 
We show that we are able to estimate the underlying glucose concentration from much smaller blood samples than is currently state-of-the-art with sufficient 
accuracy according to the most recent ISO standards and reduce measurement time significantly compared to state-of-the-art methods.  \end{abstract}
\ifCLASSOPTIONpeerreview
\begin{center} \bfseries EDICS Category: 3-BBND \end{center}
\fi
\IEEEpeerreviewmaketitle
\section{Introduction}\label{sec:intro}

\begin{figure}[hbt]\psfrag{Time (s)}[bc]{\scriptsize Time $(s)$}\psfrag{tD}[bc]{\scriptsize $n_D$}\psfrag{0}[bc]{\tiny $0$}\psfrag{t}[bc]{\scriptsize $n$}
\psfrag{RC}[l]{\scriptsize $r_C$}\psfrag{D}[l]{\scriptsize $r_D$}
\psfrag{Stage1}[tc]{\scriptsize 1)}
\psfrag{Stage2}[tc]{\scriptsize 2)}
\psfrag{Stage3}[tc]{\scriptsize 3)}
\psfrag{High Glucose Level}[bl]{\scriptsize High glucose level}
\psfrag{Low Glucose Level}[bl]{\scriptsize Low glucose level}
\psfrag{C}[bc]{\tiny $C$}\psfrag{t_C}[bc]{\tiny $n_C$}\psfrag{R}[br]{\scriptsize $r$ $(\%)$}
\psfrag{Low Glucose Concentration}[bl]{\tiny Low Glucose Level} \psfrag{High Glucose Concentration}[bl]{\tiny High Glucose Level}
\psfrag{0}[bl]{\scriptsize 0} \psfrag{5}[bl]{\scriptsize 5} \psfrag{10}[bl]{\scriptsize 10} \psfrag{15}[bl]{\scriptsize 15}\psfrag{20}[bl]{\scriptsize 20} \psfrag{50}[bc]{\scriptsize 50} \psfrag{60}[bc]{\scriptsize 60}
\psfrag{70}[bc]{\scriptsize 70} \psfrag{80}[bc]{\scriptsize 80} \psfrag{90}[bc]{\scriptsize 90} \psfrag{100}[bc]{\scriptsize 100}
\centerline{\includegraphics[width=.9\linewidth]{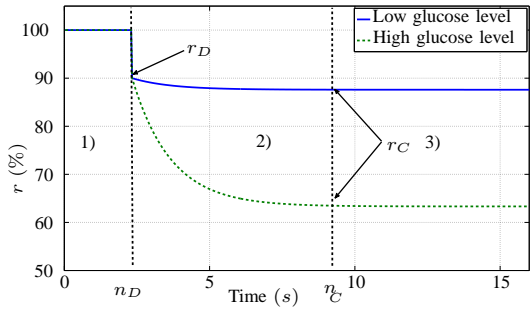}}
\vspace{-3pt}
\caption{The typical course of relative remission $r$ over time, termed kinetic curve, for a low (blue solid) and a high (green dashed) glucose level. The three distinct stages of the chemical reaction are illustrated.}
\label{fig:kinetik}
\end{figure}

\begin{figure}[hbt]
\psfrag{R}[c]{\scriptsize Intensity Value $(\%)$}
\psfrag{100}[cl]{\tiny $100$}\psfrag{0}[bc]{\tiny $0$}\psfrag{80}[cl]{\tiny $80$}\psfrag{60}[cl]{\tiny $60$}\psfrag{40}[cl]{\tiny $40$}
\psfrag{20}[cl]{\tiny $20$}\psfrag{500}[bc]{\tiny $500$}
\psfrag{1000}[bc]{\tiny $1000$}\psfrag{1500}[bc]{\tiny $1500$}
\psfrag{2000}[bc]{\tiny $2000$}\psfrag{1200}[bc]{\tiny $1200$}
\psfrag{800}[bc]{\tiny $800$}\psfrag{600}[bc]{\tiny $600$}
\psfrag{400}[bc]{\tiny $400$}\psfrag{200}[bc]{\tiny $200$}
\psfrag{30}[bl]{\tiny $30$}\psfrag{50}[bl]{\tiny $50$}
\psfrag{110}[bl]{\tiny $110$}
\psfrag{90}[bl]{\tiny $90$}\psfrag{70}[bl]{\tiny $70$}
\psfrag{x}[ct]{\scriptsize Intensity Value $(\%)$}\psfrag{y}[cc]{\scriptsize Occurrence}
\psfrag{title}[tl]{}
\begin{minipage}[b]{0.3\linewidth}
\centering
\centerline{\includegraphics[width=1\columnwidth, trim = 15cm 0cm 15cm 0cm, clip = true]{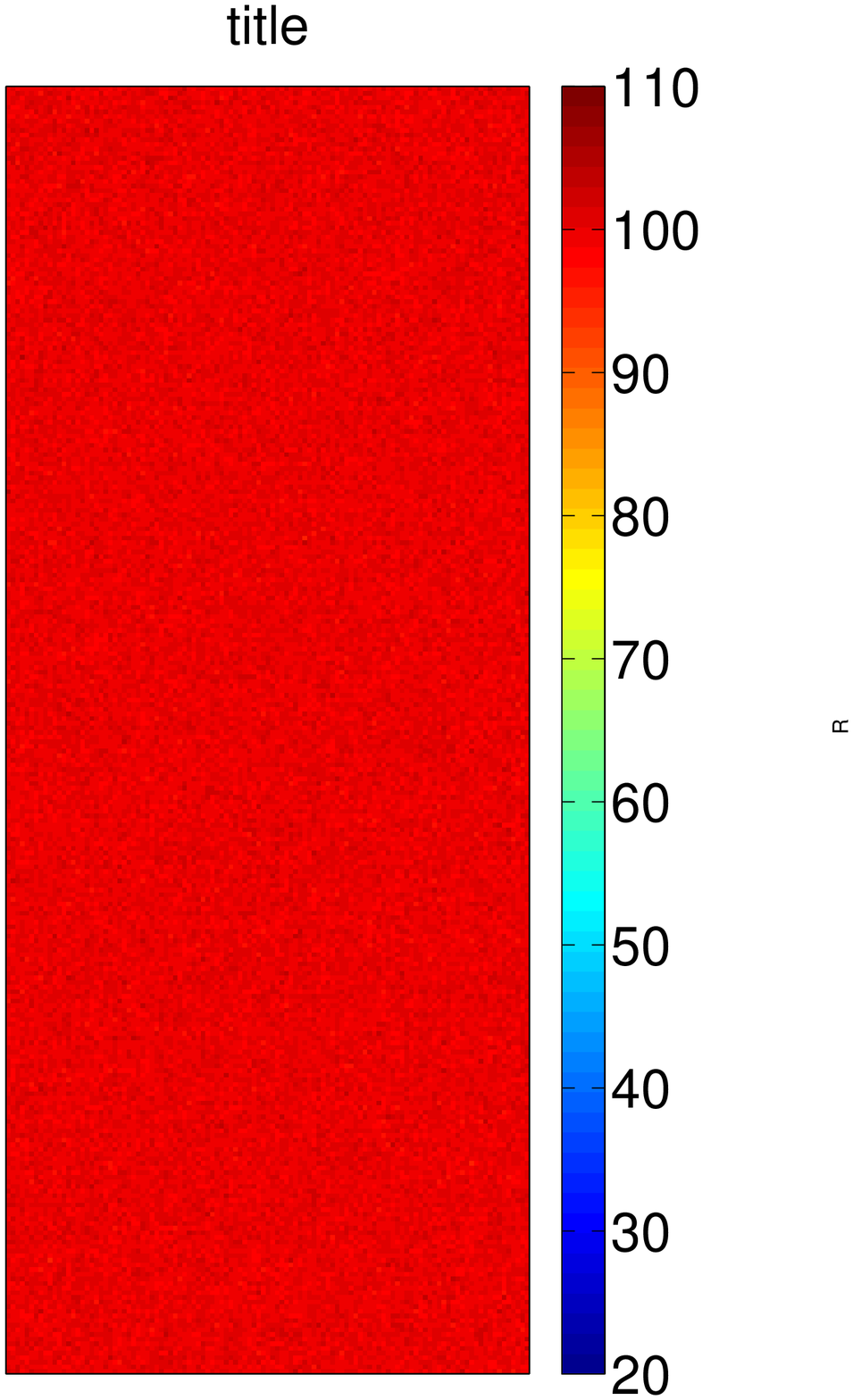}}
\vspace{-8pt}
\centerline{(a)}\medskip
\end{minipage}
\hfill
\begin{minipage}[b]{0.3\linewidth}
\centering
\centerline{\includegraphics[width=1\columnwidth, trim = 15cm 0cm 15cm 0cm, clip = true]{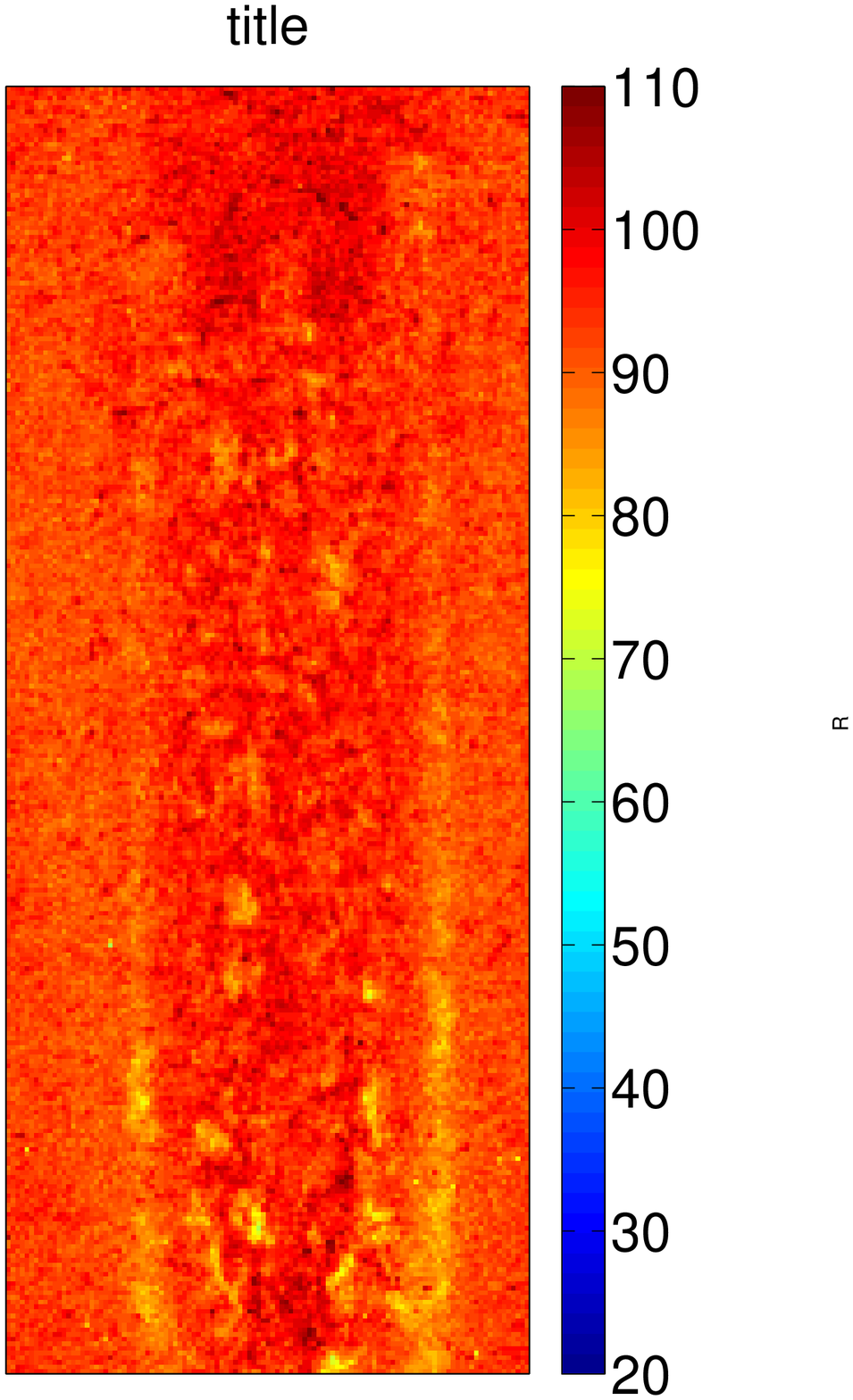}}
\vspace{-8pt}
\centerline{(b)}\medskip
\end{minipage}
\hfill
\begin{minipage}[b]{0.3\linewidth}
\centering
\centerline{\includegraphics[width=1\columnwidth, trim = 15cm 0cm 15cm 0cm, clip = true]{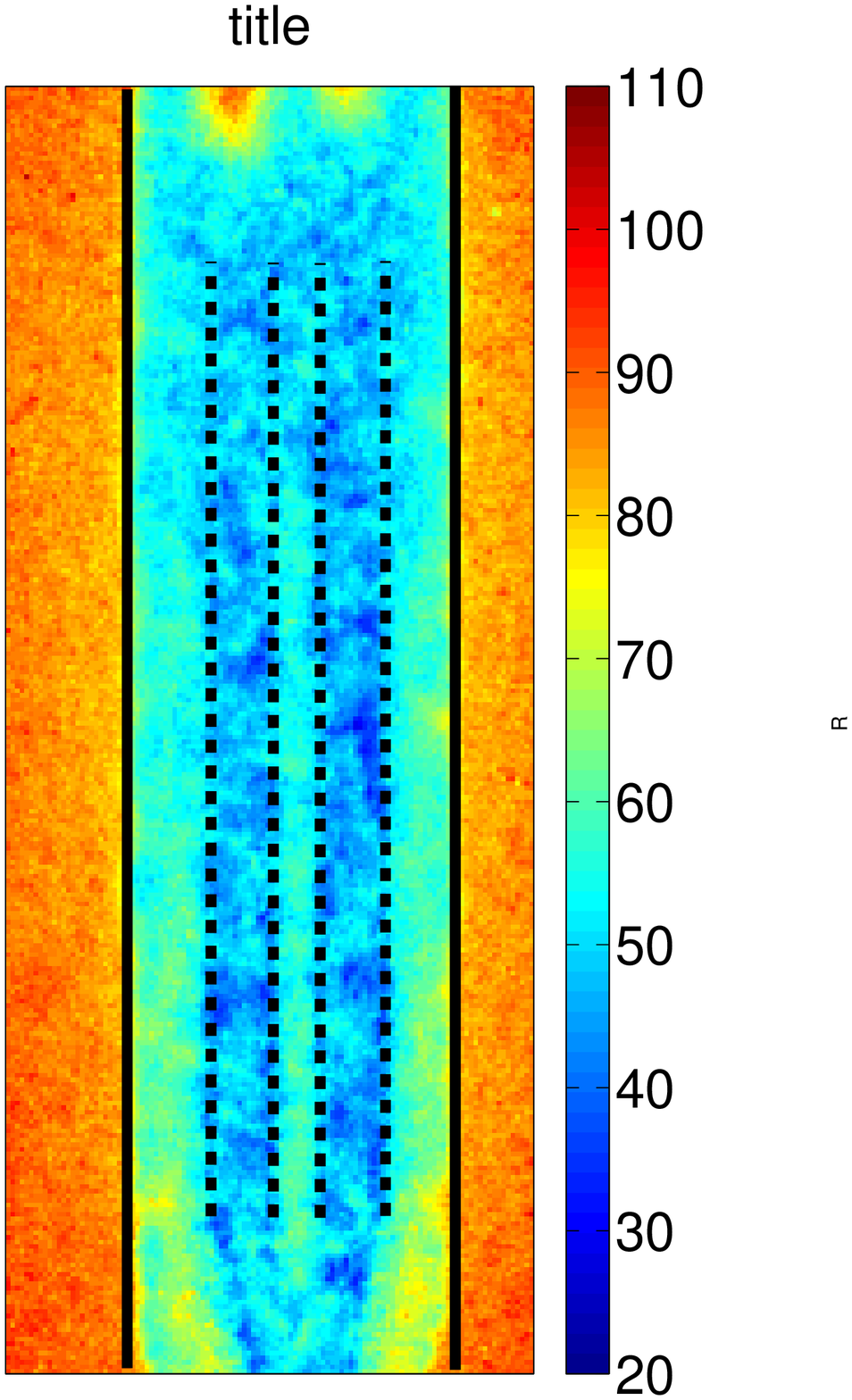}}
\vspace{-8pt}
\centerline{(c)}\medskip
\end{minipage}
\hfill
\begin{minipage}[b]{.3\linewidth}
\centering
\centerline{\includegraphics[width=1\linewidth]{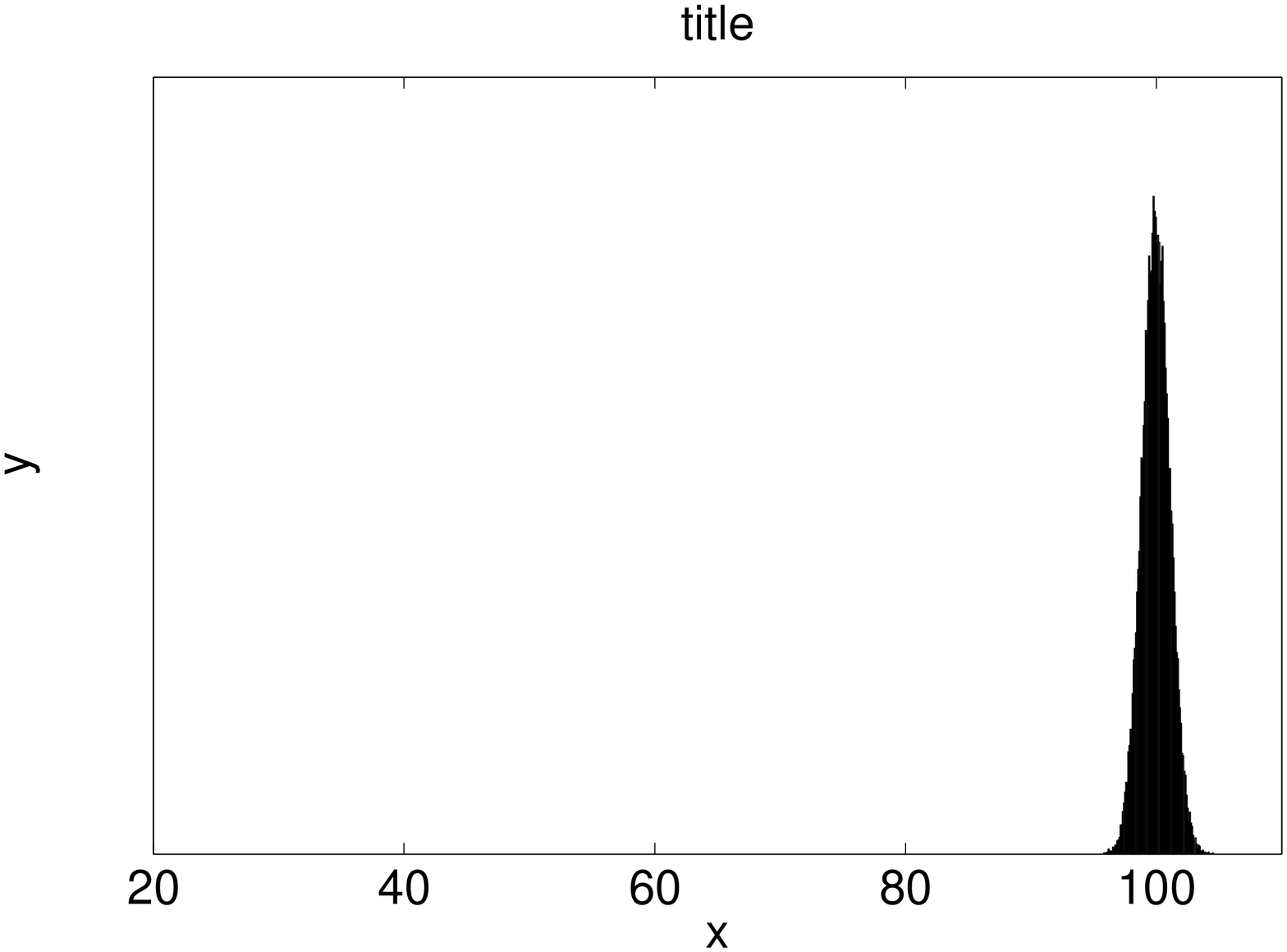}}
\vspace{8pt}
\centerline{(d)}\medskip
\end{minipage}
\hfill
\begin{minipage}[b]{0.3\linewidth}
\centering
\centerline{\includegraphics[width=1\linewidth]{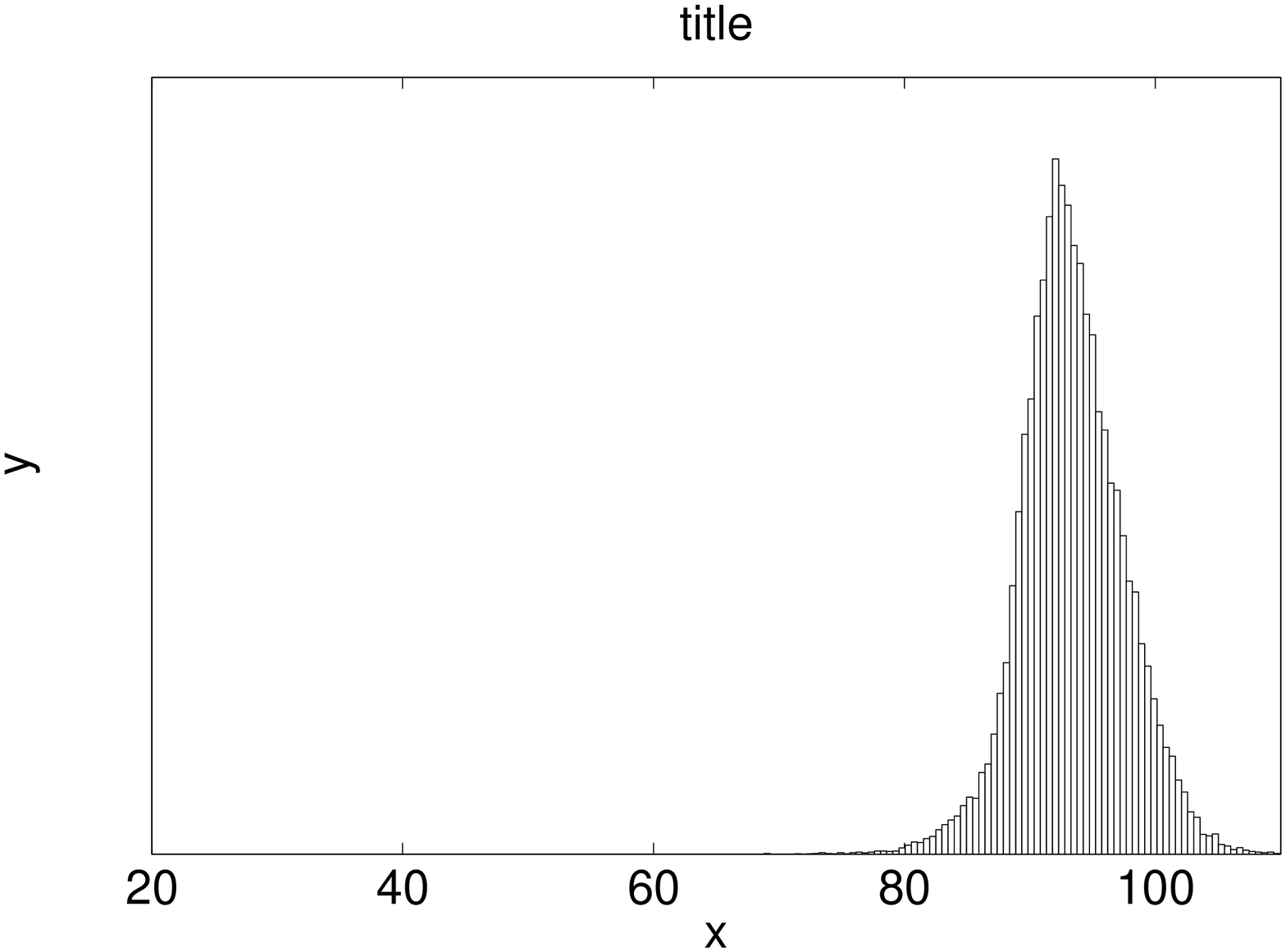}}
\vspace{8pt}
\centerline{(e)}\medskip
\end{minipage}
\hfill
\begin{minipage}[b]{.3\linewidth}
\centering
\centerline{\includegraphics[width=1\linewidth]{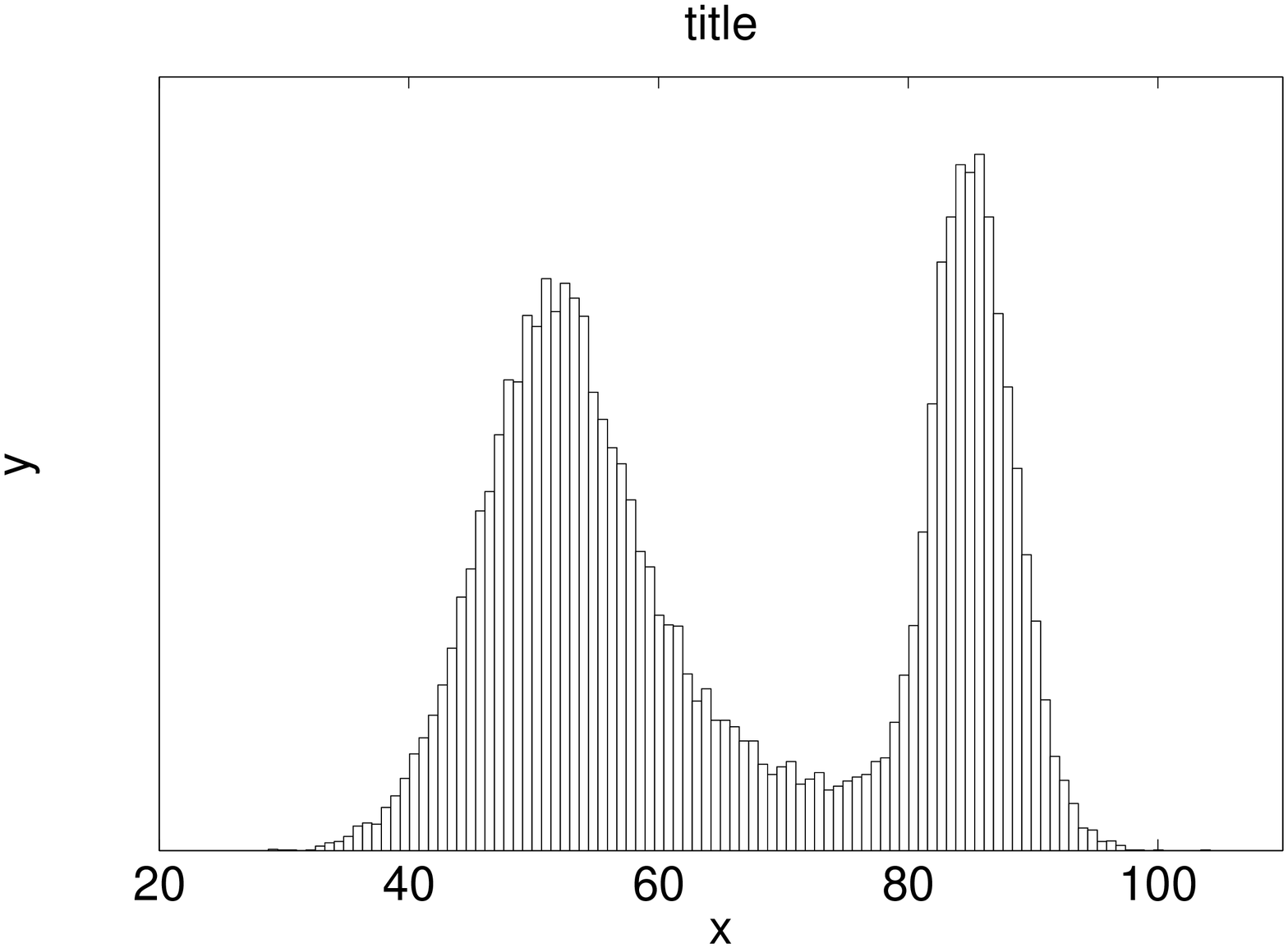}}
\vspace{8pt}
\centerline{(f)}\medskip
\end{minipage}
\hfill
\caption{Examples of observed images and their respective histograms. (a) and (d) show an observation at $n<n_D$, (b) and (e) show an observation of a low glucose measurement at $n>n_C$, (c) and (f) show an observation of a high glucose measurement at $n>n_C$.}
\label{fig:exdata}
\end{figure}

\IEEEPARstart{D}{iabetes Mellitus} describes a group of metabolic diseases, affecting 347 million people worldwide. It occurs when the pancreas cannot produce enough insulin or when the body cannot use the insulin it produces \cite{WHO}.
The healthy glucose range lies between $\SIrange{70}{180}{mg/dl}$\cite{diabetes}. A condition termed hypoglycaemia occurs when the blood sugar level drops
below $\SI{70}{mg/dl}$. This condition is associated with a high short-term risk. Hyperglycaemia, in contrast, occurs for blood sugar levels above $\SI{200}{mg/dl}$  and has longer-term effects on the body functions. Several studies have been able to statistically associate a significant delay of the onset, or slowing down the progression of
complications through intensive treatment guided by frequent blood glucose self-monitoring \cite{Control1993,Guerci2003}. For this purpose, hand-held invasive devices containing glucose biosensors are used by patients to regularly and reliably
self-monitor their glucose levels. This typically requires to extract a blood sample from the patient's finger using a lancet up to $4 - 5$ times daily and can, therefore, present a hurdle to regular self-control.
To reduce the inhibition and pain threshold for the patient, we present a framework based on a novel approach that uses a blood sample in the nano litre-range (nl-range), which is up to a magnitude of 100 smaller than is common in current state-of-the-art devices \cite{Asfour2011, spectrum2002}, hereby drastically reducing the pain sensation for the patient.
\begin{figure}[hbt]
\psfrag{R}[bc]{\tiny $R$}
\psfrag{G}[bc]{\tiny $G$}
\centerline{\includegraphics[width=1\linewidth]{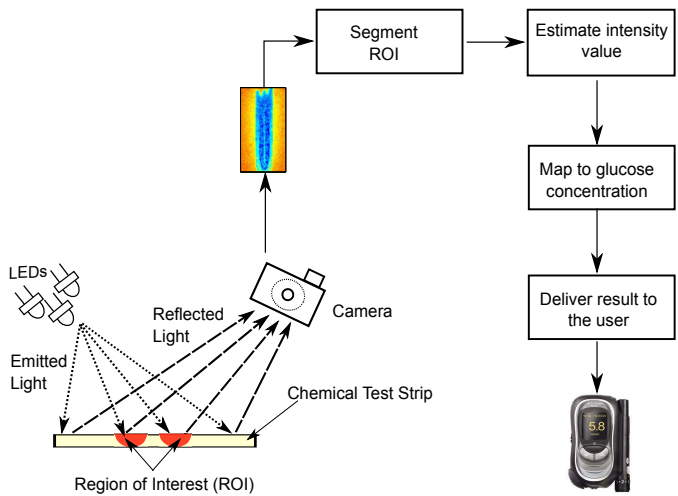}}
\vspace{-3pt}
\caption{The photometric measurement principle used to measure the glucose concentration in a blood sample.}
\label{fig:onestep}
\end{figure}
\\ The approach used in our work, illustrated in Fig.~\ref{fig:onestep}, is based on a photometric measurement principle, where the blood sample is placed on a chemical test strip that reacts with the blood glucose, resulting in a color change. 
By illuminating the test area and capturing the reflections, the color change can be measured and associated with the underlying glucose level. To counter the common problem of ambient light noise \cite{Leier09} in photometry, the measurement area is placed completely inside the device, and thus protected from ambient light. In this approach the blood sample, and thus the region where the reaction takes place, is very small compared to the chemical test strip. The
whole test strip is observed by an image sensor resulting in frames at discrete time instants $n = n_0, n_1, \ldots $ where $n = t f_s$, $t\in \mathbb{R}$ being the continuous time and $f_s$ is the sampling frequency. The frames show both the region where the reaction takes place as well as surrounding areas. The former represents the region of interest (ROI) and has to be extracted. The underlying intensity value representing the color change is termed relative remission $r \in \mathbb{R}$, which is, finally, mapped to the underlying glucose concentration $g \in \mathbb{R}$. 
This process is carried out for the whole duration of the chemical reaction, producing a set of frames.
Typically, the chemical reaction exhibits three different stages:
\begin{enumerate}
\item Constant relative remission where the reaction between the glucose and the chemical agent has not started.
\item The moistening period starts at $n = n_D$ and is characterised by a rapid drop of the relative remission value $r_D$ followed by a slow decrease, which can be modelled by an exponential decay.
\item Convergence is reached when the chemical reaction saturates at $n=n_C$ at a converged remission value $r_C$.
\end{enumerate}
Figure~\ref{fig:kinetik} depicts an idealised course of a typical chemical reaction for a low and a high glucose case. 
Different behaviours of the frames can be observed depending on the stage and the underlying glucose concentration. Figures~\ref{fig:exdata}\,(a) and \ref{fig:exdata}\,(d) show that frames captured before the chemical reaction starts are characterised by
a constant intensity over the whole region. With the onset of the chemical reaction, the ROI starts to shift in the direction of lower intensity. Depending on the underlying glucose concentration, the intensity will converge to
a different final relative remission value $r_C$. For the low glucose range (Figs.~\ref{fig:exdata}\,(b) and \ref{fig:exdata}\,(e)), this can be very close to the initial reflectance behaviour of the constant stage ($n<n_D$), whereas for high glucose concentrations (Figs.~\ref{fig:exdata}\,(c) and ~\ref{fig:exdata}\,(f)) at least three distinct areas can typically be identified.
The ROI itself is best recognisable as the area between the dotted lines in Fig.~\ref{fig:exdata}\,(c). It is characterised by a granular structure. 
Furthermore, we observe a thick edge between the ROI and the background that takes on values that lie in between both areas. In Fig.~\ref{fig:exdata}\,(c), it is the area between the solid and the dotted line. This occurs due to the inhomogeneous distribution of the blood sample over the edges, such that a weaker reaction takes place in this area.
As the images are often degraded by noise, it can be difficult to distinguish the ROI from the other image regions, particularly in low-contrast cases. The position of the ROI is unknown, as it depends on the blood flow over the test strip, as well as movements of the test strip in the camera field. 
Moreover, the ROI is often disturbed by artefacts such as air bubbles or dust particles that can change their position, size, and shape during the reaction. The result of this is a change in the number and position of clusters in the image over time as well as for different glucose concentrations. 
\\The main contribution of this work is the development of a framework to measure the concentration of an analyte in a fluid from an image-based measurement procedure; in our case the glucose concentration in blood. It is noteworthy, that this framework is not confined to this sole application but is suitable for other medical applications that rely on photometry. We show that our framework exhibits high accuracy, while reducing the blood sample volume from the micro litre-range ($\mu$l-range) to the nl-range. The proposed methodology entails: (i) the detection of the onset of the chemical reaction, using a Neyman-Pearson hypothesis test;
(ii) the segmentation of the region of interest and accurate estimation of the underlying relative remission value for each incoming frame at time instant ${n=n_D,\ldots,n_C}$ without prior knowledge on the number or shape of the regions. We build up on our previous work in \cite{DemitriSeptember2013, Demitri14}, by incorporating the median instead of the mean which exhibits better performance, as well as, providing convergence proofs for all the derived methods. 
Furthermore, a data-driven heuristic is introduced to choose the subset size individually for the scalable version, which further reduces the subset size in most cases; (iii) 
the derivation of a model for the chemical kinetics that is incorporated in a temporal tracking and prediction setup to enhance accuracy and reduce measurement time drastically; (iv) the mapping of the estimated remission to the underlying glucose concentration. 
The validation of our proposed framework is performed on an extensive collection of real data sets taken in six different scenarios to ensure robustness of the framework in different settings. Using these sets, we are able to identify a minimum range for the blood sample volume to maintain the required accuracy. We note that some aspects of this work appeared in \cite{DemitriSeptember2013,Demitri14} and \cite{Demitri15}, while the substantial part of this manuscript is novel. 
\\ Some notation: In this work, scalars are represented by lower case, non-bold letters $a$, while lower case bold letters denote vectors $\boldsymbol{a}$ and $a_l$ indicates the $l$-th element of the vector. Capital bold letters are matrices $\boldsymbol{A}$ and constants are given by capital non-bold letters $A$. We denote estimates by $\hat{a}$ and sets by $\mathcal{A}$. The use of $n$ in the superscript denotes time indices $a^{(n)}$ and $j$ in the superscript denotes iteration steps $a^{(j)}$.
\\ The paper is organised as follows: In Section \ref{sec:sota} we discuss the state-of-the-art. Section \ref{sec:proposed} presents the proposed framework, detailing the different challenges and stages of the procedure. It includes proofs of convergence of the proposed weighted mean-shift and medoid-shift algorithms. Section \ref{sec:experiments} introduces the used data sets as well 
as the validation criteria. The main part of this section comprises the experiments and the discussion thereof. 
We conclude with a summary and an overview on future work in Section \ref{sec:conclusion}.
\section{State-of-the-art}\label{sec:sota}
Traditionally, glucose self-monitoring devices are based on an invasive procedure that uses a photometric or an electrochemical approach to infer the glucose concentration from the blood sample \cite{Clarke2012, spectrum2002, Shin2003}. This requires the extraction of a blood volume of $\SIrange{1}{25}{\micro l}$. 
While much research is being performed to replace the current devices by non-invasive alternatives \cite{Kayashima91, Yamaguchi98, Cameron97}, it seems that the traditional technologies will continue to maintain their position\cite{spectrum2002}, at least in the near future. Reducing the blood volume needed, and hereby the induced pain remains a necessary field of research \cite{spectrum2002}. Traditional photometric measurement principles use a blood sample in the $\mu$l-range that completely covers up the test field. A photodiode is applied to capture the resulting reflections \cite{Baumann1995}. The use of much smaller blood samples renders this approach inaccurate, as the ROI becomes much smaller and the signal-to-noise-ratio rises. This motivates the use of a camera to observe the chemical reaction.
To the best of our knowledge, the problem of measuring blood glucose using the setup in Fig.\ref{fig:onestep} and an image processing-based approach has only been tackled in a limited number of studies. In \cite{Asfour2011}, the authors propose to use a histogram-based approach and the setup in Fig.~\ref{fig:onestep} to
estimate the intensity of the region of interest. To this end, two clusters are assumed, one corresponding to the ROI and the other to the background. The authors propose to measure the displacement of the ROI cluster w.r.t. the background cluster. The underlying assumptions are that just two clusters exist in the image and that the background cluster exhibits a constant relative remission value over time. 
These are rather strong assumptions, which are hardly fulfilled in practice. Our observations give evidence to a variable number of clusters in the images, as seen in Fig.~\ref{fig:exdata}. 
The variability is glucose level-specific, temporally dependent, and caused by the variability of the chemical reaction for different blood samples of different patients. Furthermore, the background cluster does not experience a constant intensity but changes over time 
due to temperature and humidity issues, as well as leakage of small amounts of blood to the background cluster.
We assume that the number of clusters in the images is unknown and propose an approach that is able to deal systematically with this.
\section{Proposed Algorithm}
\label{sec:proposed}
\begin{figure}[hbt]
\psfrag{n}[b]{\footnotesize $n$}
\psfrag{I}[b]{\footnotesize $\Ibf^{(n)}$}
\psfrag{G}[b]{\footnotesize $\hat{g}$}
\psfrag{R}[b]{\footnotesize $\hat{r}_C$}
\psfrag{n = n+1}[l]{\footnotesize $n = n+1$}
\psfrag{Preprocessing}[c]{\footnotesize Pre-processing}
\centerline{\includegraphics[width=1\linewidth]{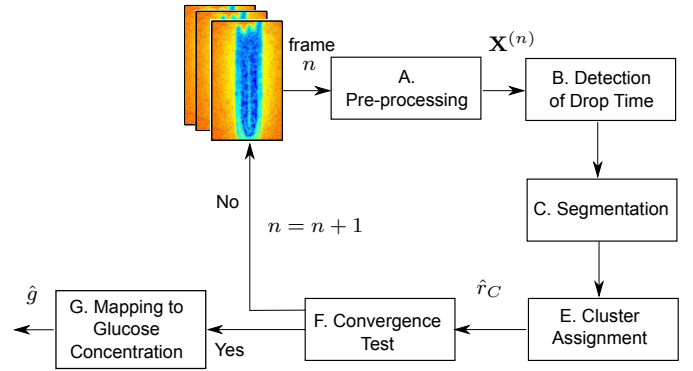}}
\vspace{-3pt}
\caption{The proposed approach.}
\label{fig:fg_prop_1}
\end{figure}

Our framework is summarised in Fig.~\ref{fig:fg_prop_1}. The input is given by the frames obtained by the camera, representing the reflectivity behaviour of the observed area. Each incoming frame is directly processed.
A hypothesis test is performed on the pre-processed, gray-scale reflectance image $\Ibf^{(n)}$ of size $M_x\times M_y$, obtained at frame $n$, to detect whether the chemical reaction has started, i.e., if $n\geq n_D$. In this case, we proceed by segmenting the region of interest.  
After segmenting the image, the relative remission values of the different regions are estimated and the converged relative remission value $\hat{r}_C$ corresponding to the ROI is identified. The current estimate along with the history of estimates is used to test for convergence of the chemical reaction. If convergence is detected, the underlying intensity value is mapped to its corresponding glucose level $\hat{g}$. 
If not, the next frame is processed in the same manner. 
The different blocks as depicted in Fig.~\ref{fig:fg_prop_1} will be described in further detail in the sequel.  
\subsection{Pre-processing} 
The raw frames $\Ibf^{(n)}_\text{raw}$ obtained by the camera are normalised w.r.t. the initial reflectivity prior to the start of the chemical reaction. 
The normalisation is performed using the calibration frames as described in \cite{DemitriSeptember2013} and results in a gray-level image $\Ibf^{(n)}_{\text{norm}}$ with pixels in the range between $0-100$.
\\In our previous work \cite{DemitriSeptember2013} we asserted that windowing the images results in superior segmentation and remission estimation results. We termed this process binning and used $B$ to denote the size of the window. 
Binning is used as the last step in the pre-processing stage unless stated otherwise. Finally, the pre-processed image is denoted by $\Ibf^{(n)}$.
 \subsection{Detection of Drop Time}\label{sec:detectdrop}
\begin{figure}[h]
\psfrag{x}[tt]{\scriptsize $\tilde{\x}$}
\psfrag{y}[cc]{\scriptsize $f_{\tilde{\x}}$}
\psfrag{t1}[ll]{\scriptsize $n_1$}
\psfrag{t2}[ll]{\scriptsize $n_2$}
\psfrag{-30}[tt]{\scriptsize -30}\psfrag{-20}[tt]{\scriptsize -20}  \psfrag{-10}[tt]{\scriptsize -10} \psfrag{10}[tt]{\scriptsize 10} \psfrag{20}[tt]{\scriptsize 20} \psfrag{30}[tt]{\scriptsize 30} \psfrag{40}[tt]{\scriptsize 40}
\psfrag{50}[tt]{\scriptsize 50} \psfrag{95}[bc]{\scriptsize 95} \psfrag{105}[bc]{\scriptsize 105} \psfrag{115}[bc]{\scriptsize 115}
\psfrag{0}[tr]{\scriptsize } \psfrag{0.05}[cr]{\scriptsize 0.05} \psfrag{0.15}[cr]{\scriptsize 0.15}\psfrag{0.25}[cr]{\scriptsize 0.25}\psfrag{0.35}[cr]{\scriptsize 0.35}\psfrag{0.4}[cr]{\scriptsize 0.4}
\psfrag{0.1}[cr]{\scriptsize 0.1} \psfrag{0.2}[cr]{\scriptsize 0.2}\psfrag{0.3}[cr]{\scriptsize 0.3}
\psfrag{title}[bc]{} 
\centerline{\includegraphics[width=.6\linewidth]{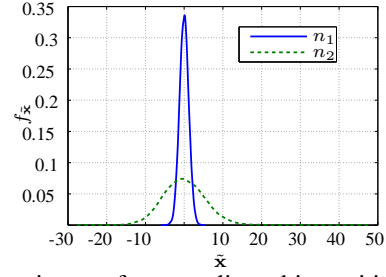}}
\caption{Pdf estimate of mean-adjusted intensities of a frame at $n_1 < n_D$ (blue solid) and $n_2 \geq n_D$ (green dashed).}
\label{fig:NP_dropdet}
\end{figure}

As illustrated in Fig.~\ref{fig:kinetik}, the chemical reaction exhibits a constant stage for $n<n_D$, where the glucose in the blood sample has not yet been detected. The time $n_D$ when the drop occurs, i.e., when the chemical reaction starts, is dependent on the particular chemical reagent, the blood sample at hand, and the underlying temperature and humidity conditions in the device. 
We suggest using the Neyman-Pearson hypothesis test of variance to detect $n_D$. The advantage of using the Neyman-Pearson test is that it provides a meaningful way to set a threshold by incorporating the image statistics and controlling the false-alarm rate.
Figure~\ref{fig:NP_dropdet} illustrates the probability density functions $f_{\tilde{\x}}$ of the intensities of two exemplary images taken at time $n_1<n_D$ and $n_C > n_2 >n_D$, i.e., directly after the occurrence of the drop. Extensive analysis of real data showed that it is reasonable to assume that the mean-adjusted frames for $n<n_D$ follow a zero-mean normal distribution with variance $\sigma_1^2$, whereas after the drop the mean-adjusted frames can be modelled by a zero-mean normal distribution with variance $\sigma_2^2 > \sigma_1^2$. This increase in variance can be explained by the onset of the chemical reaction and thereby the onset of the color change on the test strip. Hereby, $\sigma_1^2$ can be estimated as the sample variance from available data sets. However, $\sigma_2^2$ is unknown and depends on the underlying glucose concentration. 
\\ For the remainder of this work, we will deal with vectorised forms of the images and denote these by 
  \begin{align} \label{eq:vec} \x^{(n)} = \text{vec}\left(\Ibf^{(n)}\right), \end{align}
where $\x^{(n)}$ is of size $1\times L, L = M_x\cdot M_y$. Now, the hypothesis test can be formulated as
 \begin{align}
 \mathcal{H}_0:&  f_{\tilde{\x}^{(n)}} \sim \mathcal{N}(0,\sigma_1^2), \quad n < n_D \\
  \mathcal{H}_1:&  f_{\tilde{\x}^{(n)}} \sim \mathcal{N}(0,\sigma_2^2), \quad n \geq n_D \nonumber
 \end{align}
where $\tilde{\x}^{(n)}$ is the mean-adjusted version of $\x^{(n)}$ and $f_{\tilde{\x}^{(n)}}$ denotes the pdf of $\tilde{\x}^{(n)}$.
Calculating the likelihood-ratio leads to the following test statistic
 \begin{align}
\func{\textit{T}}{\tilde{\x}^{(n)}} = \sum^{L}_{l=1} \left(\tilde{x}^{(n)}_l\right)^2  \underset{H_0}{\overset{H_1}{\gtrless}} \underbrace{L\frac{\frac{2}{L}\ln(\delta)+ \ln(\frac{\sigma_2^2 }{\sigma_1^2 })}{\frac{1}{\sigma_1^2 }-\frac{1}{\sigma_2^2 }}}_{\delta'},
 \end{align}
where the threshold $\delta$ is set to ensure a nominal false alarm rate. Assuming $\tilde{\x}^{(n)}$ to be spatially i.i.d., the distribution of $\frac{\textit{T}(\tilde{\x}^{(n)})}{\sigma_1^2}$ under $ \mathcal{H}_0$ and $\frac{\textit{T}(\tilde{\x}^{(n)})}{\sigma_2^2}$ under $ \mathcal{H}_1$ can be shown to follow a ${\chi}^2$-distribution with $L$ degrees of freedom \cite{Kay93}. The spatial i.i.d. assumption is justifiable in our case as we are observing frames prior to the onset of the reaction where no structure is present in the image. $L$ represents the number of pixels in an image and is, therefore, quite large. 
Thus, we can approximate the ${\chi}^2$-distribution by a Gaussian distribution \cite{Kay93} and the probability of false alarm $ P_{\text{FA}}$ becomes
 \begin{align}
 P_{\text{FA}} = \func{Q}{\frac{\delta'}{\sigma_1^2}}, 
 \end{align}
 where $\func{Q}{\cdot}$ is the complementary cumulative distribution function of the standard Gaussian distribution. 
Given a fixed probability of false alarm $P_{\text{FA}}$, the threshold  can be calculated as $\delta' = \func{Q^{-1}}{P_{\text{FA}}}\cdot\sigma^2_1$.
\subsection{Segmentation Using the Mean-Shift Algorithm}
\subsubsection{The Standard Mean-Shift Algorithm}
Our underlying assumption is that the glucose images will contain a ROI, where the blood sample is distributed and, thus, the color change of the ROI represents the reflectivity caused by the underlying glucose concentration. This 
area will not necessarily be completely spatially connected as it may contain artefacts where the reaction has not taken place, or granularities in the chemical that do not contribute to a proper reaction. Other areas in the image will not correspond to the ROI, either, because the blood sample is unevenly distributed in these areas, or because the areas are dry, i.e., not covered with blood. 
We interpret the different areas contained in the image as clusters with specific cluster centres overlaid by noise. We assume that data points converging to a certain cluster center will belong to the corresponding area. The noise here is assumed to be Gaussian. This assumption is based on the calibration frames where no reaction has started. 
\\ The mean-shift algorithm \cite{Fukunaga} has become a popular approach for clustering and mode location estimation. 
It has been widely applied for medical image segmentation as it 
does not require the knowledge of the actual number of clusters \cite{DemitriSeptember2013, Comaniciu2002, Lee2005, Janowczyk2012, Debeir2005}. 
We use the mean-shift algorithm (MS), as well as two extensions of it: the robust 
mean-shift (R-MS) \cite{Demitri14}
and the scalable, sparse mean-shift (SS-MS) \cite{Demitri15}. Furthermore, we extend these variants to the medoid-shift \cite{Sheikh2007a}. 
\\ Assume a set of $L$ pixels in a gray-scale image\footnote{Including spatial information for gray-scale images will increase the dimensions by two, leading to a three-dimensional data vector ${\x = [x^{\text{intensity}}, x^{\text{position},M_x}, x^{\text{position},M_y}]}$.} at frame $n$ that can be expressed as $x_l^{(n)} \in \mathbb{R}$, $ l=1,\ldots,L$ as in Eq.~(\ref{eq:vec}).
We will omit the frame notation, henceforth, for better readability. 
\\ A consistent estimator of the density is given by the kernel density estimator (KDE) \cite{silverman} with bandwidth parameter $h$
\begin{equation}\label{eq:standardKDE}
\hat{f}_K(\x) = \frac{1}{Lh}\sum_{l = 1}^{L}K\left(\frac{\x-x_l}{h}\right),
\end{equation}
where $K(\x)$ is a radially symmetric kernel function with a strictly decreasing profile for $x\geq 0$. The bandwidth $\h$ of the kernel function is the only parameter that has to be tuned and is crucial to the performance of the mean-shift algorithm, as it affects the number of clusters.
Extensive work has dealt with the choice of the bandwidth parameter, e.g. \cite{Comaniciu2001, silverman}. 
\\ Generally, the mean-shift algorithm is derived by taking the zeros of the gradient of the KDE
\begin{align}
\nabla \hat{f}_K(\x) = 0
\end{align}
and reformulating to get the mean-shift vector of $\x$ as
\begin{align}\label{eq:msvec}
\MS =\left[\frac{\sum_{l=1}^L x_l \func{K'}{\frac{\x-x_l}{h}}}{\sum_{l=1}^L \func{K'}{\frac{\x-x_l}{h}}}-\x\right],
\end{align}
where $K' = \frac{dK}{d\x}$. Each data point is shifted according to Eq.~(\ref{eq:msvec}) in the direction of steepest ascent until convergence, such that the sequence of successive locations of a data vector
\begin{align}\label{eq:msvec2}
\x^{(j+1)} =\left[\frac{\sum_{l=1}^L x_l \func{K'}{\frac{\x^{(j)}-x_l}{h}}}{\sum_{l=1}^L\func{K'}{\frac{\x^{(j)}-x_l}{h}}}\right], \quad j = 1,2,...
\end{align}
where $\x^{(j)}$ denotes the $j$-th iteration step. 
Typically, the mean-shift algorithm tends to overestimate the number of modes. It is usual to post-process the mean-shift results with a simple clustering method to group nearby modes together, e.g. using a mode pruning step with bandwidth $h$~\cite{DemitriSeptember2013,Comaniciu2002}.
\\ Another way to express the KDE, which allows for extensions of the mean-shift, is to consider the kernel to be an inner product in the Hilbert space $\mathcal{H}$, such that\cite{Scholkopf2002a} 
\begin{align}\label{eq:KT}
K\left(\frac{\x-x_l}{h}\right) = \langle \Phi(\x),\Phi(x_l) \rangle,
 \end{align} 
 where $\Phi: \mathbb{R^d} \rightarrow \mathcal{H}$ is a mapping function and $\langle\cdot\rangle$ denotes the inner product. $K(\frac{\x-x_l}{h})$ is a positive definite kernel function. Hence, the KDE can be formulated as 
\begin{equation}\label{eq:kernel_KDE}\hat{f}(\x) =  \langle \Phi(\x),\sum_{l=1}^{L} w_l\Phi(x_l) \rangle.\end{equation}
In this formulation, the uniform weights $\frac{1}{L}$ have been substituted by $w_l>0$ to attain a more general form.
It has been shown \cite{Comaniciu2002} that the mean-shift algorithm converges and that the mode estimator is asymptotically consistent and unbiased. 
\subsubsection{The Robust Mean-Shift Algorithm}
In \cite{Demitri14}, the robust mean-shift (R-MS) is derived as an alternative to account for heavy-tailed noise in the data. Here, the sample mean of vectors in Eq.~(\ref{eq:kernel_KDE}) is substituted by a robust M-estimate
\begin{align}
{\hat{\mu}}_{\Phi} = \underset{{{\mu}_\Phi}}{\arg\min} \sum_{l = 1}^{L} \func{\rho}{\frac{\Phi(x_l)-{\mu}_{\Phi}}{\hat{\sigma}}},
\end{align}
where $\func{\rho}{\cdot}$ is a monotone, differentiable loss function, such as Huber's loss function \cite{Huber1964a} and the scale $\hat{\sigma}$ is initialized with a robust estimate based on the mean absolute deviation \cite{Demitri14}. 
This leads to the robust KDE
\hspace{-5pt}\begin{align} \label{eq:RKDE}
\hat{f}(\mathbf{x}) &= \langle\Phi(\mathbf{x}),{{\hat\mu}_{\Phi}}\rangle = \Big\langle\Phi(\mathbf{x}),\sum_{l = 1}^{L}w_l^{\text{R-MS}}\Phi(x_l)\Big\rangle \nonumber \\ 
		    &=\frac{1}{\h}\sum_{l = 1}^{L}w_l^{\text{R-MS}}K\left(\frac{\x-x_l}{h}\right),
\end{align}
where the robust weights $w_l^{\text{R-MS}}$ can be determined using Iteratively ReWeighted Least Squares (IRWLS) \cite{Demitri14,  Maronna2006a}, resulting in
\begin{align}
w_l = \left\{
\begin{array}{cl}
\frac{\hat{\sigma}\cdot\func{\psi}{{\frac{\Phi(x_l)-{\mu}_{\Phi}}{\hat{\sigma}}}}}{\Phi(x_l)-{\mu}_{\Phi}} & \text{if} \quad \frac{\Phi(x_l)-{\mu}_{\Phi}}{\hat{\sigma}} \neq 0\\
\func{\psi'}{0} & \text{if} \quad \frac{\Phi(x_l)-{\mu}_{\Phi}}{\hat{\sigma}} = 0\end{array},\right.
\end{align}
where $\psi = \rho'$.
The robust mean-shift vector reads 
\begin{align}
\RMS =\left[\frac{\sum_{l=1}^L w_l^{\text{R-MS}}x_l \func{K'}{\frac{\x-x_l}{h}}}{\sum_{l=1}^L w_l^{\text{R-MS}}\func{K'}{\frac{\x-x_l}{h}}}-\x\right].
\end{align}
For a more detailed description of the R-MS the reader is referred to \cite{Demitri14}.
\subsubsection{The Scalable Sparse Mean-Shift Algorithm}\label{sec:SS-MS}
\begin{figure}[hbt!]
\psfrag{y}[bb]{\scriptsize $\widetilde{\nu}_{\mathcal{I}}$}
\psfrag{x}[tc]{\scriptsize $N$}\psfrag{D}[bl]{\tiny $n<n_D$}
\psfrag{High}[bl]{\tiny High Glucose}\psfrag{Low}[bl]{\tiny Low Glucose}
\psfrag{100}[bc]{\scriptsize 100}\psfrag{0}[bc]{\tiny 0}\psfrag{200}[bc]{\scriptsize 200}\psfrag{400}[bc]{\scriptsize 400}\psfrag{600}[bc]{\scriptsize 600}\psfrag{800}[bc]{\scriptsize 800}\psfrag{1000}[bc]{\scriptsize 1000}\psfrag{1}[bc]{\tiny $1$}
\psfrag{1200}[bc]{\scriptsize 1200}\psfrag{10}[bc]{\tiny $10$}\psfrag{0.2}[bc]{\tiny $0.2$}\psfrag{0.4}[bc]{\tiny 0.4}\psfrag{0.6}[bc]{\tiny $0.6$}\psfrag{0.8}[bc]{\tiny $0.8$}\psfrag{0e}[lc]{\tiny $10^0$}\psfrag{1e}[lc]{\tiny $10^1$}\psfrag{2e}[lc]{\tiny $10^2$}\psfrag{3e}[lc]{\tiny $ L = 10^3$}
\psfrag{Title}[bc]{}
\psfrag{0.1}[bc]{\tiny $0.1$}\psfrag{0.3}[bc]{\tiny $0.3$}\psfrag{0.5}[bc]{\tiny}\psfrag{0.7}[bc]{\tiny $0.7$}\psfrag{0.9}[bc]{\tiny $0.9$}
\centerline{\includegraphics[width=.75\linewidth]{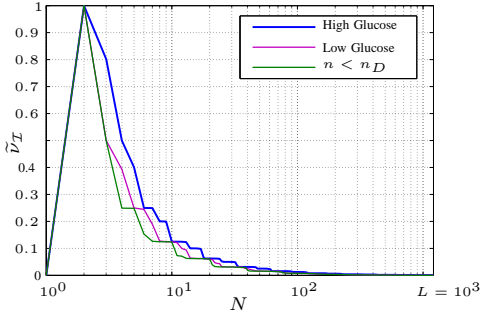}}
\vspace{-3pt}
\caption{An example of the progression of normalised $\nu$ with increasing cardinality $N$.}
\label{fig:nu_over_N}
\end{figure}

The computational complexity of the mean-shift algorithm is proportional to the square of the total number of data points $L$. To overcome the problem of high computational demand, a scalable sparse version has been derived in \cite{Demitri15}. A more detailed discussion of the computational complexity of both versions is provided in \cite{Demitri15}. Using only a subset of the data points, the SS-MS is able to achieve high accuracy while reducing computational power. 
The gist lies in substituting the mean in Eq.~(\ref{eq:kernel_KDE}) by a sparse approximation\cite{Scott2014,Cortes2015}. A short outline will be given in the sequel, for more details the reader is referred to \cite{Demitri15}.
The sparse approximation of the mean can be formulated as
\begin{equation}
\min_{\mathcal{I}_{|\mathcal{I}| = N}}\min_{\alpha_i, i \in \mathcal{I} }\left|\left| \sum_{l = 1}^{L}w_l\langle \Phi(\x),\Phi(x_l) \rangle - \sum_{i \in \mathcal{I} }\alpha_i\langle \Phi(\x),\Phi(x_i) \rangle\right|\right|^2. 
\label{eq:sparseapprox2}\end{equation}
where the index set $\mathcal{I} \subseteq \{1,\ldots,L\}$ of cardinality $|\mathcal{I}| = N$ is defined to be a subset of the full set of indices, the weights follow $\boldsymbol{\alpha} \in \mathbb{R}^N$ and $1\leq N \ll L$.
The scalable, sparse KDE 
\begin{align}\label{eq:SSKDE}
\hat{f}_K(\x) = \frac{1}{\h}\sum_{i\in\I^*}\alpha_{\mathcal{I}^*,i}K\left(\frac{\x-x_i}{h}\right),\end{align}
where $\alpha_{\mathcal{I}^*,i}$ denotes the $i$-th weight obtained using the optimal set $\mathcal{I}^*$.
The resulting mean-shift vector has the form
\begin{align}
\SSMS =\left[\frac{\sum_{i\in\I^*} \alpha_{\mathcal{I}^*,i}x_i \func{K'}{\frac{\x-x_i}{h}}}{\sum_{i\in\I^*} \alpha_{\mathcal{I}^*,i}\func{K'}{\frac{\x-x_i}{h}}}-\x\right]. \end{align}
A robust sparse formulation (RSS-MS) can be derived \cite{Demitri15}.
\\To solve Eq.~(\ref{eq:sparseapprox2}), we need to: 1) find a solution for the inner optimisation problem; i.e., solve for $\boldsymbol{\alpha}$: 2) find an optimal set $\mathcal{I}^*$.
For a fixed $ \mathcal{I} $, $ \boldsymbol{\alpha}_{\mathcal{I}} = \boldsymbol{\Xi}^{-1}_{ \mathcal{I}}\boldsymbol{\xi}_{ \mathcal{I}}$. Here 
\begin{align} 
\boldsymbol{\Xi}_{ \mathcal{I}} &= (\langle \Phi(x_i), \Phi(x_j)\rangle)_{i,j \in \mathcal{I}}\\
 \boldsymbol{\xi}_{ \mathcal{I}} &= \sum_{j = 1}^{L}w_j\langle \Phi(x_m), \Phi(x_j)\rangle, {m \in \mathcal{I}},
\end{align}
where $\boldsymbol{\Xi}_{ \mathcal{I}} $ is the Gram matrix. To ensure convergence, ${\alpha_{i}<0, \forall i = 1,...,N}$. Generally, $\alpha_{i}$ will not always fulfill the property. We, therefore, normalise the weights $\alpha_{i}$ by setting to zero all weights $\alpha_{i}<0$ and re-normalising such that $\sum_i\alpha_{i}=1$.
The optimal index subset is found by assuming a fixed $N$ and maximising an incoherence function $\nu_{\mathcal{I}}$
\begin{align}\label{eq:nu}
\mathcal{I}^* = \max_{\small{\mathcal{I} \subseteq \{1,\ldots,L\}}}\nu_{\mathcal{I}}, \end{align}
where 
\begin{align}
\nu_{\mathcal{I}} = \min_{j\notin\mathcal{I}}\max_{i \in \I} \langle \Phi(x_i), \Phi(x_j)\rangle.
\end{align}
This is intuitive in the sense that to find the most representative subset of data vectors, we choose the ones that are most incoherent to each other. 
In~\cite{Demitri15}, we outline an algorithm to maximize $\nu_{\mathcal{I}}$ and determine $\mathcal{I}^*$.
The question remaining is how to choose the cardinality $N$ of the subset of indices $\mathcal{I}$. Clearly, in some cases this can be given by the application at hand knowing a certain $N_\text{max}$, or by using
test data for cross-validation, as in \cite{Demitri15}, where $N$ is set to be sufficiently high to incorporate the worst-case scenario. We introduce a data-driven selection of $N$, which results in a smaller $N$ for most data sets than cross-validation. 
Figure~\ref{fig:nu_over_N} shows typical examples of the progression of a normalised version of $\nu_{\mathcal{I}}$ 
\begin{align}
\widetilde{\nu}_{\mathcal{I}} = \frac{\nu_{\mathcal{I},|\mathcal{I}|=N}}{\max\nu_{\mathcal{I}}}
\end{align}
for the images given in Fig.~\ref{fig:exdata}, using different values of $N = 1, \ldots, L$. 
We observe that the value of $\widetilde{\nu}_{\mathcal{I}}$ drops quickly after a certain value of $N$ and that the behaviour of the curve is similar for different images. 
This signifies that the first $N_\nu$ samples chosen by the algorithm contribute highly to the incoherence in the image and it is sufficient to use a subset $\mathcal{I}$ of cardinality $N_{\nu}$ for the sparse representation. 
We propose to choose $N_\nu$ based on the gradient of $\widetilde{\nu}_{\mathcal{I}}$ in Fig.~\ref{fig:nu_over_N} and set a threshold $T_{\nu}$
\begin{align}
\Delta \widetilde{\nu}^{(N_\nu)} = \frac{\nu_{\mathcal{I}, |\mathcal{I}| = N_\nu}}{\nu_{\text{max}}} - \frac{\nu_{\mathcal{I}, |\mathcal{I}| = N_\nu-1}}{\nu_{\text{max}}} \leq T_{\nu}.
\end{align}
The choice of $T_{\nu}$ is based on heuristics and chosen specifically for the data at hand by observing the trade-off between accuracy and computation. 
\subsubsection{Convergence Properties of the Derived Algorithms}
It remains to prove that the mean-shift algorithm in all its variations will converge. In \cite{Comaniciu2002}, the convergence proof of the standard MS is given. We extend it for the cases of the R-MS and the SS-MS.
\begin{mydef}
If  $w_l>0$ holds for the weights from Eq.~(\ref{eq:kernel_KDE})  and the kernel function $K(\cdot)$ has a convex and monotonically decreasing profile, the sequence of trajectory points $\{\x^{(j)}\}$, ${j = 1,2,\ldots}$ will converge and the sequence $\{\func{\hat{f}_K}{\x^{(j)}}\}$, ${j = 1,2,\ldots}$ is monotonically increasing. 
\end{mydef}
The proof given in Appendix \ref{sec:app1} is a generalised version of {Theorem 1} in \cite{Comaniciu2002}. We need to ensure that the weights will fulfill $w_l^{\text{R-MS}}>0$ for R-MS and $\alpha_{\I^*,i}>0$ for SS-MS. $w_l^{\text{R-MS}}$ are derived using the IRWLS and are, thus, certain to follow the required property for monotone loss functions $\func{\rho}{{\cdot}}$. The calculation of $\alpha_{\I^*,i}$ ensured that it follows this property. 
\subsection{Segmentation Using the Medoid-Shift Algorithm}
The medoid-shift was derived by Sheikh \textit{et al.}~in \cite{Sheikh2007a} as an alternative to the mean-shift that uses the medoid instead of the mean. The medoid is defined as the point in the set that has the smallest distance to all other points. 
We derive here a robust medoid-shift and a sparse scalable medoid-shift. This results in the following formulation for successive locations of $\x$:
\begin{align}\label{eq:msvec2_med}
\x^{(j+1)} =  \underset{\x \in \{x_l\}}{\arg\min} \sum_{l=1}^L \Big|\Big|\x-x_l\Big|\Big|^2w_l\func{K'}{\frac{\x^{(j)}-x_l}{h}}, 
\end{align} 
where $w_l$ represents the weights associated with each kernel. For the standard medoid shift (MedS) $w_l = \frac{1}{L}$ and for the robust medoid-shift (R-MedS) $w_l =  w_l^{\text{R-MS}}$. For the sparse scalable medoid-shift \\ 
\begin{align}\label{eq:ssmsvec2_med}
\x^{(j+1),\text{SS-MedS}} =  \underset{\x \in \{\x_i\}}{\arg\min} &\sum_{i\in \mathcal{I}^*} \Big|\Big|\x-x_i\Big|\Big|^2 \times\\\nonumber 
&\alpha_{\mathcal{I}^*,i}\func{K'}{\frac{\x^{(j)}-x_i}{h}}
\end{align} 
Unlike the mean-shift, the medoid-shift always converges to points contained in the data set. The advantage of this is that the medoid-shift vector needs only to be computed once for every data sample and hence, needs less iterations than the mean-shift. Moreover, it does not need a further heuristic like the mean-shift to group together neighbouring clusters.
The convergence of the sequence of trajectory points in the medoid-shift is guaranteed and the proof is analogous to that of the mean-shift. The only difference is that for the medoid-shift we additionally need to ensure that there are no cycles in the sequence of trajectory points, i.e., $\x^{(j)}\neq\x^{(j+c)}$, for all $c>0$. The proof thereof is given in Appendix \ref{sec:app_med}.
\subsection{Cluster Assignment}
The segmentation leaves us with a finite number of cluster centres corresponding to the different image regions. Typically, we attain $2 - 4$ clusters; 1) one corresponding to the ROI, 2) another to the dry test strip area and optionally 3)
one to the border between the dry area and the ROI, and 4) one corresponding to artefacts in the image. We need to identify which cluster center corresponds to the ROI. 
This can be done using a data-driven approach, exploiting information from the structure of the images. Using the size of the clusters, we can identify the background region and the ROI, as they are the largest. The ROI will correspond to the one exhibiting the lower intensity of the two remaining regions according to the nature of the chemical reaction.
\subsection{Convergence Test}\label{sec:EKFconvergence}
\begin{figure}[hbt]
\psfrag{y}[c]{\scriptsize $\hat{\tau}$}
\psfrag{x}[bc]{\scriptsize $\hat{r}_C$}
\psfrag{fitted curve}[bl]{\scriptsize LS fit}
\psfrag{data points}[bl]{\scriptsize Data points}
\psfrag{70}[bc]{\scriptsize 70} \psfrag{80}[bc]{\scriptsize 80} \psfrag{90}[bc]{\scriptsize 90} \psfrag{100}[bc]{\scriptsize 100} \psfrag{50}[bc]{\scriptsize 50} \psfrag{60}[bc]{\scriptsize 60}
\psfrag{0}[bc]{\scriptsize 0}
\psfrag{-0.05}[bl]{\scriptsize -0.05}\psfrag{-0.1}[bl]{\scriptsize -0.1}\psfrag{-0.2}[bl]{\scriptsize -0.2}\psfrag{-0.3}[bl]{\scriptsize -0.3}\psfrag{-0.4}[bl]{\scriptsize -0.4}\psfrag{-0.15}[bl]{\scriptsize -0.15}\psfrag{-0.25}[bl]{\scriptsize -0.25}\psfrag{-0.35}[bl]{\scriptsize -0.35}
\psfrag{title}[bc]{}
\centerline{\includegraphics[width=.8\linewidth]{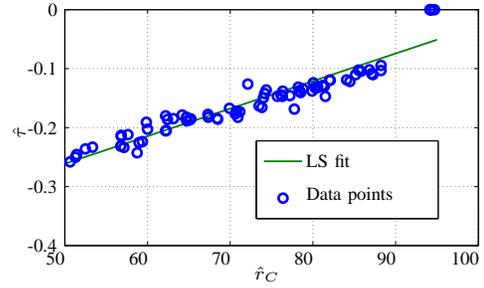}}
\caption{Regression analysis for the relation between $\tau$ and $r_C$}
\label{fig:kalmantau}
\end{figure}

Convergence of the chemical reaction is a very crucial issue for the estimation accuracy. Ideally, the chemical kinetic curve evolves as in Fig.~\ref{fig:kinetik}. Final convergence is typically reached after $\SIrange{10}{15}{s}$. This leads to a long measurement time for the patient and, hereby, reduced usability, which can be a cause of irregular self-control. 
So far, the convergence of kinetic curves in hand-held devices has been defined as the point at which the slope of the kinetic curve reaches a threshold $T_\text{slope}$ over a predefined time period \cite{Asfour2011}. This point is reached after around $\SIrange{3}{15}{s}$, depending on the underlying measurement. 
The assumption made, hereby, is that in the ideal case, measurements of equal glucose concentration will follow a similar progression at all times and, therefore, the intensity value estimate at the predefined slope threshold will be related to the underlying glucose concentration. 
The mapping function between intensity and glucose can be adjusted to account for the inaccuracy. While this can lead to satisfactory results when no information is given on the kinetic curve, we will show that it can also result in erroneous, premature estimates.
We propose to improve the performance of this approach drastically by incorporating a model of the kinetic behaviour of the chemical reaction~\cite{Leier09}, along with state estimation techniques to predict the actual convergence values ahead of time. Hereby, stages 2) and 3) of the kinetic curve depicted in Fig.~\ref{fig:kinetik} are modelled.
\\ The glucose oxidase taking place on the chemical test strip can be modelled by the differential equation~\cite{Leier09}
\begin{align}\label{eq:model1}
r(n) &= (r_D - r_C) \cdot e^{-n\tau} + r_C + v(n), \quad n>n_D,
\end{align}
where $r(n)$ is the remission measurement value at time $n$, $r_0$ is the initial remission value after the drop, and $r_C$ is the convergence value of the relative remission after the chemical reaction has converged. We define $r_C$ to be the state that we want to predict. 
Furthermore, $\tau$ is the reaction rate, and $v(n)$ is a zero-mean, white, Gaussian noise process with unknown variance $\sigma_v^2$ describing the measurement noise. 
\\ Due to the nature of the chemical reaction, it is realistic to assume that $\tau$ and $r_C$ are correlated. Building up on ~\cite{Leier09}, we perform a regression analysis using a real data set of estimated convergence values $\mathbf{\hat{r}}_{C} = [\hat{r}_{C,1},...,\hat{r}_{C,N_M}]$ and corresponding rates $\boldsymbol{\hat{\tau}} = [\hat{\tau}_{1},...,\hat{\tau}_{N_M}]$, $N_M$ being the number of available measurements. Hereby, we establish a linear relation between $\tau$ and $r_C$ to be the most suitable least-squares (LS) fit, as can be seen in Fig.~\ref{fig:kalmantau}
\begin{align}
\boldsymbol{\hat{\tau}} = \Delta\tau \cdot \mathbf{\hat{r}}_{C} + \tau_0, 
\end{align}
where $\Delta\tau > 0$ and $\tau_0 < 0$. 
Using a least-squares (LS) fit, we estimate the regression parameters $\Delta\tau$ and $\tau_0$, to obtain from Eq.~(\ref{eq:model1}) the following nonlinear relation for the kinetic curve
\begin{align}\label{eq:model2}
r(n) &= (r_D - r_C) \cdot e^{-n\cdot (\Delta\tau \cdot \mathbf{r}_{C} + \tau_0)} + r_C + v(n) \\\ &= \func{f}{r_C,v(n)} ,  \quad n>n_D \nonumber.
\end{align}
We use an Extended Kalman Filter (EKF) to process the measured kinetic curve and perform an online prediction of the remission convergence value $r_C$. For that, we formulate the prediction and measurement equations as follows
\begin{align}
  \hat{r}_C(n) &= {a}(n) \hat{r}_C(n-1) + w(n) \\
 \hat{r}(n) &= \func{h}{\hat{r}_C(n),v(n)}  , 
\end{align}
where ${a}(n)$ is the transition matrix describing the transition of the state estimate $\hat{r}_C$ from time $n-1$ to $n$. In our case, ${{a}(n) = 1}$, as $r_C$ is a static state of the chemical reaction. The process noise is modelled by the random variable $w(n)$ and can be used to account for uncertainty in the model. $\func{h}{\hat{r}_C(n),v(n)}$ describes the observation model, relating the estimated state $\hat{r}_C(n)$ to the measurement $\hat{r}(n)$. It is given by
the partial derivative of the process model w.r.t. the state $\hat{r}_C$
\begin{align}
\func{h}{\hat{r}_C(n),v(n)} &= \frac{\partial f(\hat{r}_C(n),V(n))}{\partial \hat{r}_C(n)} \\ \nonumber 
                & = 1 - e^{-n\cdot (\Delta\tau\cdot \hat{r}_C(n) + \tau_0)} \\
                &\times \left(1 + r_D \cdot \Delta\tau \cdot n - \hat{r}_C(n) \cdot \Delta\tau \cdot n\right). \nonumber
\end{align}
We alternate between the prediction and the correction step of the EKF, including a new measurement $\hat{r}(n)$ in each iteration.
\vspace{-15pt}
\subsection{Mapping to Glucose Concentration}\label{sec:codecurve_sec}
Finally, we map the relative remission estimate $\hat{r}_C$ to the underlying glucose concentration, which we will deliver to the user. 
To this end, a calibration function ${f_\text{Calib}\,:\,\hat{r}_C\rightarrow\,\hat{g}}$ is coded into the glucometer to perform the mapping during measurement. 
The calibration function has to be generated \textit{a priori} by means of photometric measurements performed under lab conditions, using blood samples with known underlying glucose concentrations. 
\section{Experimental Results and Discussion}\label{sec:experiments}
\subsection{Data Sets}\label{sec:datasets}
\begin{table}[htb]
\centering\caption{The real data sets used for validation.}
\rowcolors{2}{gray!25}{white}
\small\begin{tabular}{ c c c c c}
\rowcolor{gray!50}
\toprule
Set & $N_M$ & $N_g$& $\Upsilon$& Volume\\\midrule
A & 48 & 5 & 6.45 $\mu m /$ Pixel& \SIrange{10}{100}{nl}\\\hline 
B & 78 & 16 & 6.45 $\mu m /$ Pixel& \SIrange{10}{100}{nl}\\\hline 
C & 78 & 16 & 6.45 $\mu m /$ Pixel& \SIrange{10}{100}{nl}\\\hline 
D & 48 & 4 & 30 $\mu m /$  Pixel& \SIrange{10}{100}{nl}\\ \hline 
E & 200 & 10 & 30 $\mu m /$ Pixel& around $\SI{1}{nl}$\\ \hline 
F & 200 & 10 & 30 $\mu m /$ Pixel& Standard\\ \bottomrule 
\end{tabular}\vspace{10pt}
\label{tab:datasets}\end{table}
To evaluate our proposed framework, we use real data sets, obtained from a setup as in Fig.~\ref{fig:onestep} using blood samples from blood donations injected with a glucose solution
corresponding to the amount of glucose needed. Altogether, we have six different data sets, comprising 452 different measurements which corresponds to a total of 263.064 processed images. 
The sets differ in terms of the chemical used as well as the resolution of the camera and the volume of the blood sample. This results in a different blood flow over the test strip and thereby different positions and shapes of the region of interest, as well as prominence of the reaction. 
Information on the different data sets is given in Table \ref{tab:datasets}. The number of measurements in the data sets is denoted by $N_M$,
the number of different glucose concentrations tested in each data set by $N_g$, and the resolution of the images by $\Upsilon$. The volume of the blood drop "Standard" indicates state-of-the-art ranges around $\SIrange{1}{25}{\micro l}$\cite{spectrum2002}. Each measurement contains $N_{f} = 580$ frames obtained at a frame rate of $f_s = \SI{30}{fps}$. 
All data sets are pre-processed applying a binning size of $B = 5$ for $\Upsilon = \SI{6.45}{\micro m}$ /Pixel and $B = 1$ for $\Upsilon = \SI{30}{\micro m}$/Pixel. After binning, all images contain $L = 1210$ pixels. For the KDE, a Gaussian kernel and a fixed bandwidth parameter $\h$ as in \cite{DemitriSeptember2013} are employed. $T_{\text{slope}}$ is chosen to be $10^{-2}$. For the robust versions, as in \cite{Demitri14}, Huber's loss function is used and its parameter tuned to achieve 95$\%$ asymptotic efficiency in the Gaussian case; IRWLS is initialised with uniform weights.
\vspace{-10pt}
\subsection{Validation Methods}
\subsubsection{Coefficient of Variation}
The coefficient of variation (CV) is a very popular measure to assess the accuracy of pharmacokinetic measurements \cite{CVcitation}. In our work, we use the remission coefficient of variation $\text{CV}_{\hat{r}}$
\begin{equation}
\text{CV}_{\hat{r}} = \frac{1}{N_g}\sum_{\gamma = 1}^{N_g} \text{CV}_{\hat{r}_{g(\gamma)}},\quad\text{CV}_{\hat{r}_{g(\gamma)}} = \frac{\hat{\sigma}_{\mathcal{\hat{R}}_{g(\gamma)}}}{\hat{\mu}_{\mathcal{\hat{R}}_{g(\gamma)}}}\end{equation}
where $\gamma = 1,\ldots,N_g$, $\hat{\sigma}_{\mathcal{\hat{R}}_{g(\gamma)}}$ and $\hat{\mu}_{\mathcal{R}_{g(\gamma)}}$ are the sample standard deviation, and respectively the sample mean of the elements of the set $\mathcal{\hat{R}}_{g(\gamma)}$.
$\mathcal{\hat{R}}_{g(\gamma)}$ denotes the test set of relative remission estimates obtained from several measurements of the same underlying glucose concentration $g(\gamma)$. 
\subsubsection{The Clarke Error Grid}
The Clarke Error Grid (CEG) \cite{delfavero12} is a standard method for evaluating glucose measurement performance. 
It plots the estimated glucose concentrations against the actual glucose concentrations and classifies the error according to its medical severity. To this end, different regions are defined in the CEG as depicted in Fig.~\ref{fig:CEG_results}. 
A common specification is to have at least $95\%$ of all points in the region A, maximally $5\%$ of all points in the region B \cite{CEG2} and no points in the other zones.
\subsubsection{The Glucose-Specific Mean Absolute Deviation}
To quantify the validation, we employ a glucose specific mean absolute deviation (MAD)  inspired by \cite{delfavero12}.
Instead of using the standard MAD between the true and the estimated glucose level, 
a glucose-specific MAD (gMAD) is defined. The gMAD weighs the errors using a penalty function according to their medical severity on the basis of the CEG. 
Hereby, errors made for hypoglycaemic cases below $\SI{70}{mg/dl}$ are weighed with the highest factor $w_{\text{gMAD}}$ as they present critical short-term risks; hyperglycaemic cases which present more long-term risks are weighed with a slightly smaller factor $w_{\text{gMAD}}$, and normal cases are given the factor $w_{\text{gMAD}} = 1$. The gMAD reads
\begin{equation}
\text{gMAD} = \frac{1}{N} \sum_{\gamma = 1}^{N_M} \left|g(\gamma) - \hat{g(\gamma)}\right|\cdot w_{\text{gMAD}}(g(\gamma) - \hat{g(\gamma)}),
\end{equation}
where the penalty function is given by\begin{equation}
w_{\text{gMAD}}(g - \hat{g})= \left\{ \begin{array}{ll}
         1.5 \overline{\sigma}(g) \sigma(\hat{g}) & \mbox{if $g \leq 85 $ and $ \hat{g}\geq g$};\\
         1 \sigma(g) \overline{\sigma}(\hat{g}) & \mbox{if $g \geq 155 $ and $ \hat{g}\leq g$};\\
        1 & \mbox{otherwise},\end{array} \right.
\end{equation}
with $\overline{\sigma}(\cdot)$ and ${\sigma}(\cdot)$ being sigmoid functions that ensure a smooth transition as given in \cite{delfavero12}. According to the most recent ISO standards~\cite{ISO2013}, the maximal permissible error is $\pm \SI{15}{mg/dl}$ within a reference range of $\SIrange{0}{75}{mg/dl}$ and $20\%$ for a reference range above $\SI{75}{mg/dl}$. 
\subsection{Results}
First, we analyze the quality of the remission results. Herein, a comparison of the segmentation methods is given, as well as an analysis of the different data sets. Next, we evaluate the data-driven choice of $N_\nu$ for the SS-MS. We, then, turn our attention to the effect of using the EKF to predict the convergence estimates. Finally, we study the precision of the glucose estimates after the mapping operation.
\subsubsection{Kinetic Curves \& Remission Accuracy}
\begin{figure}[hbt!]
\psfrag{200}[bl]{\scriptsize 200}\psfrag{300}[bl]{\scriptsize 300}\psfrag{400}[bl]{\scriptsize 400}\psfrag{500}[bl]{\scriptsize 500}
\psfrag{40}[bl]{\scriptsize 40} \psfrag{0}[bl]{} \psfrag{5}[bl]{\scriptsize 5} \psfrag{10}[bl]{\scriptsize 10} \psfrag{15}[bl]{\scriptsize 15} \psfrag{70}[bl]{\scriptsize 70} \psfrag{80}[bl]{\scriptsize 80} \psfrag{90}[bl]{\scriptsize 90} \psfrag{100}[bl]{\scriptsize 100} \psfrag{50}[bl]{\scriptsize 50} \psfrag{60}[bl]{\scriptsize 60}
\psfrag{140}[bl]{\scriptsize 140}\psfrag{130}[bl]{\scriptsize 130}\psfrag{120}[bl]{\scriptsize 120}\psfrag{110}[bl]{\scriptsize 110}\psfrag{MedS}[bl]{}\psfrag{RMedS}[bl]{}\psfrag{RSMS}[bl]{}\psfrag{MS}[bl]{}
\begin{minipage}[b]{0.49\linewidth}
\centering
\psfrag{y}[rc]{\scriptsize $\hat{r}$ $(\%)$}
\psfrag{Glc}[c]{\scriptsize Glucose Level [mg/dl]}
\psfrag{x}[bc]{\scriptsize Time [s]}
\psfrag{title}[bc]{}
\centerline{\includegraphics[width=1\linewidth]{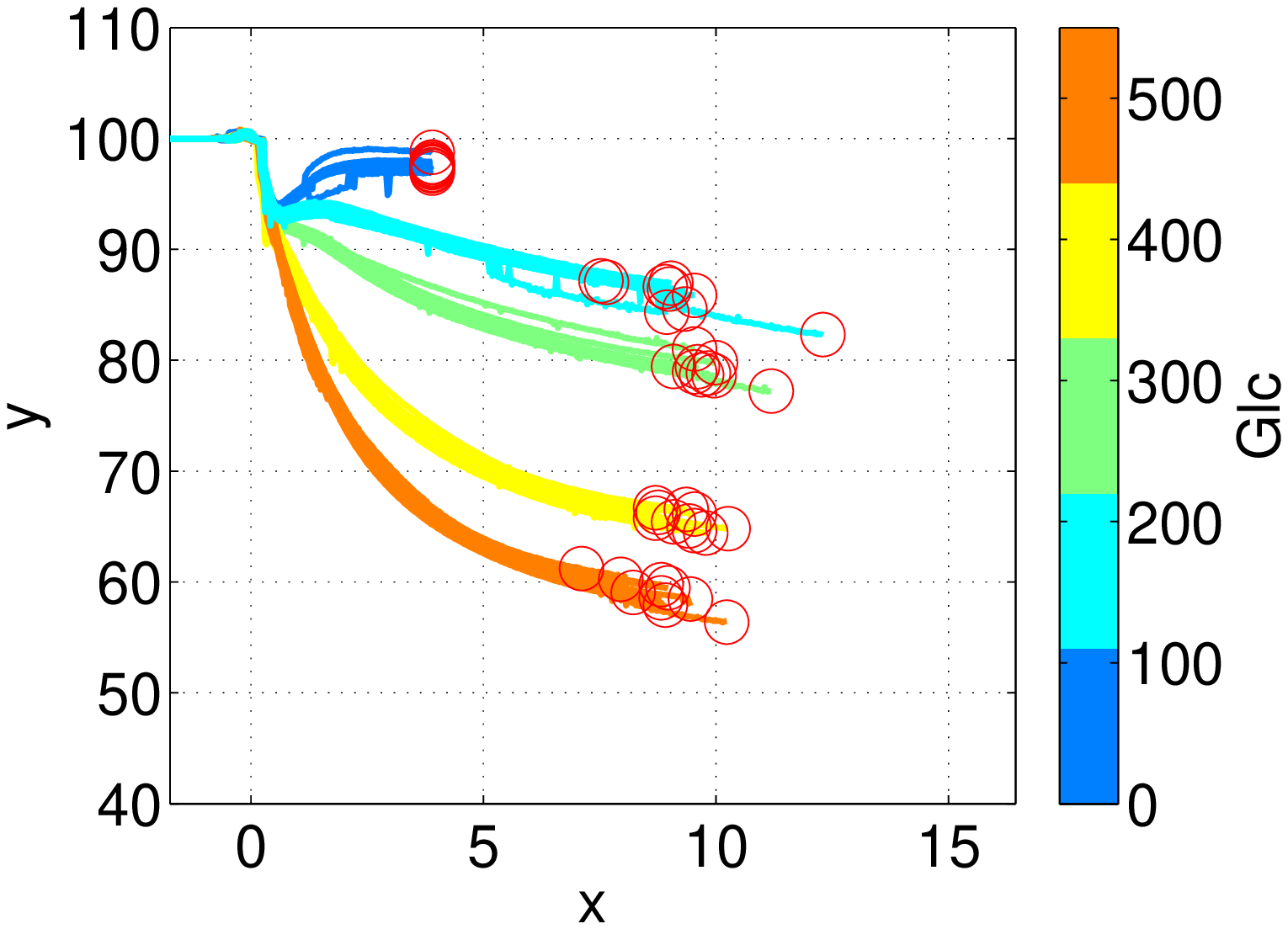}}
\vspace{-3pt}
\centerline{(a) Set A using MS}\medskip
\end{minipage}
\hfill
\begin{minipage}[b]{0.49\linewidth}
\centering
\psfrag{y}[rc]{\scriptsize $\hat{r}$ $(\%)$}
\psfrag{Glc}[c]{\scriptsize Glucose Level [mg/dl]}
\psfrag{x}[bc]{\scriptsize Time [s]}
\psfrag{title}[bc]{}
\centerline{\includegraphics[width=1\linewidth]{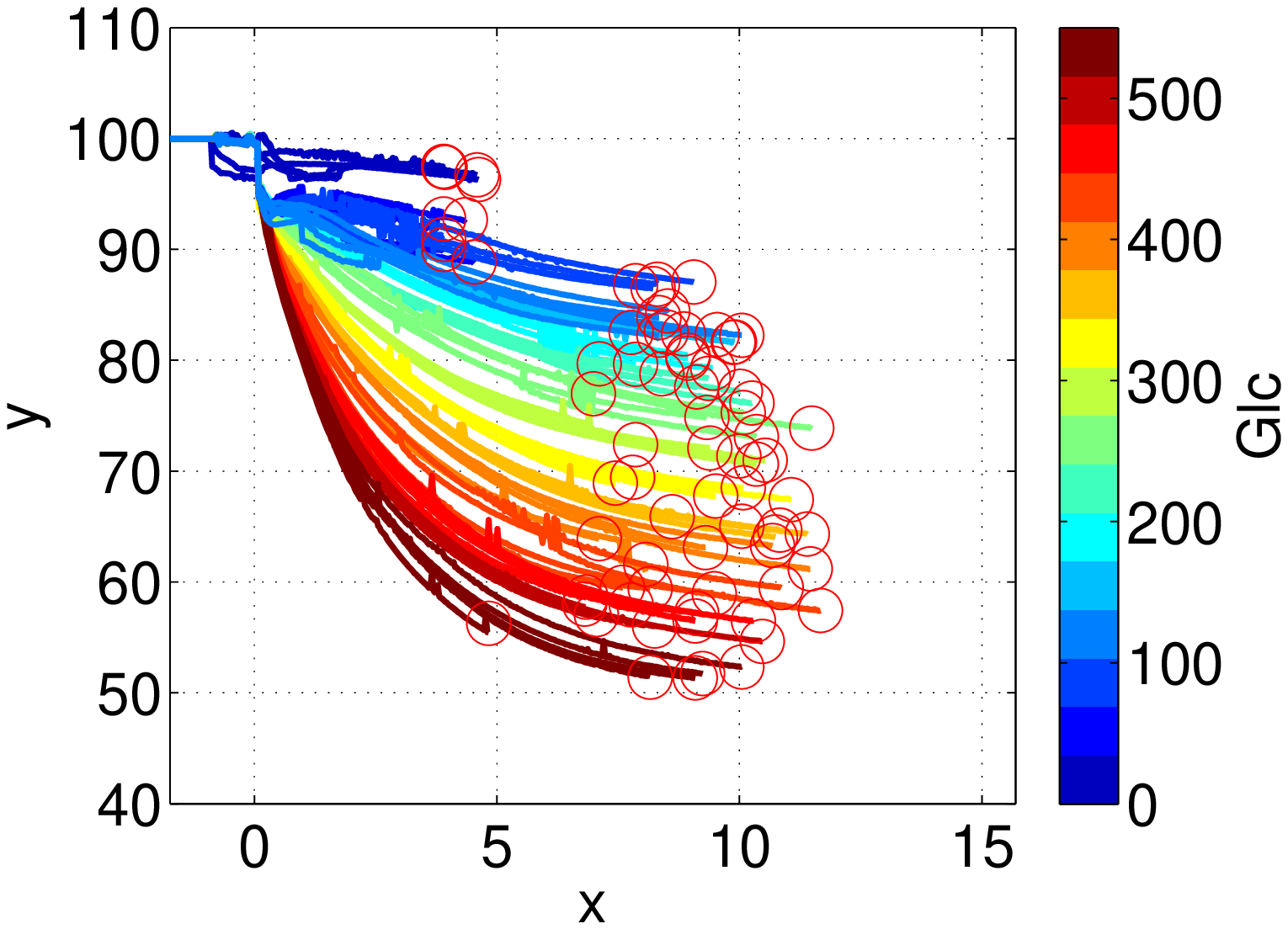}}
\vspace{-3pt}
\centerline{(b) Set C using RSS-MS}\medskip
\end{minipage}
\hfill
\begin{minipage}[b]{.49\linewidth}
\centering
\psfrag{y}[rc]{\scriptsize $\hat{r}$ $(\%)$}
\psfrag{x}[bc]{\scriptsize Time [s]}
\psfrag{Glc}[c]{\scriptsize Glucose Level [mg/dl]}
\psfrag{title}[bc]{}
\centerline{\includegraphics[width=1\linewidth]{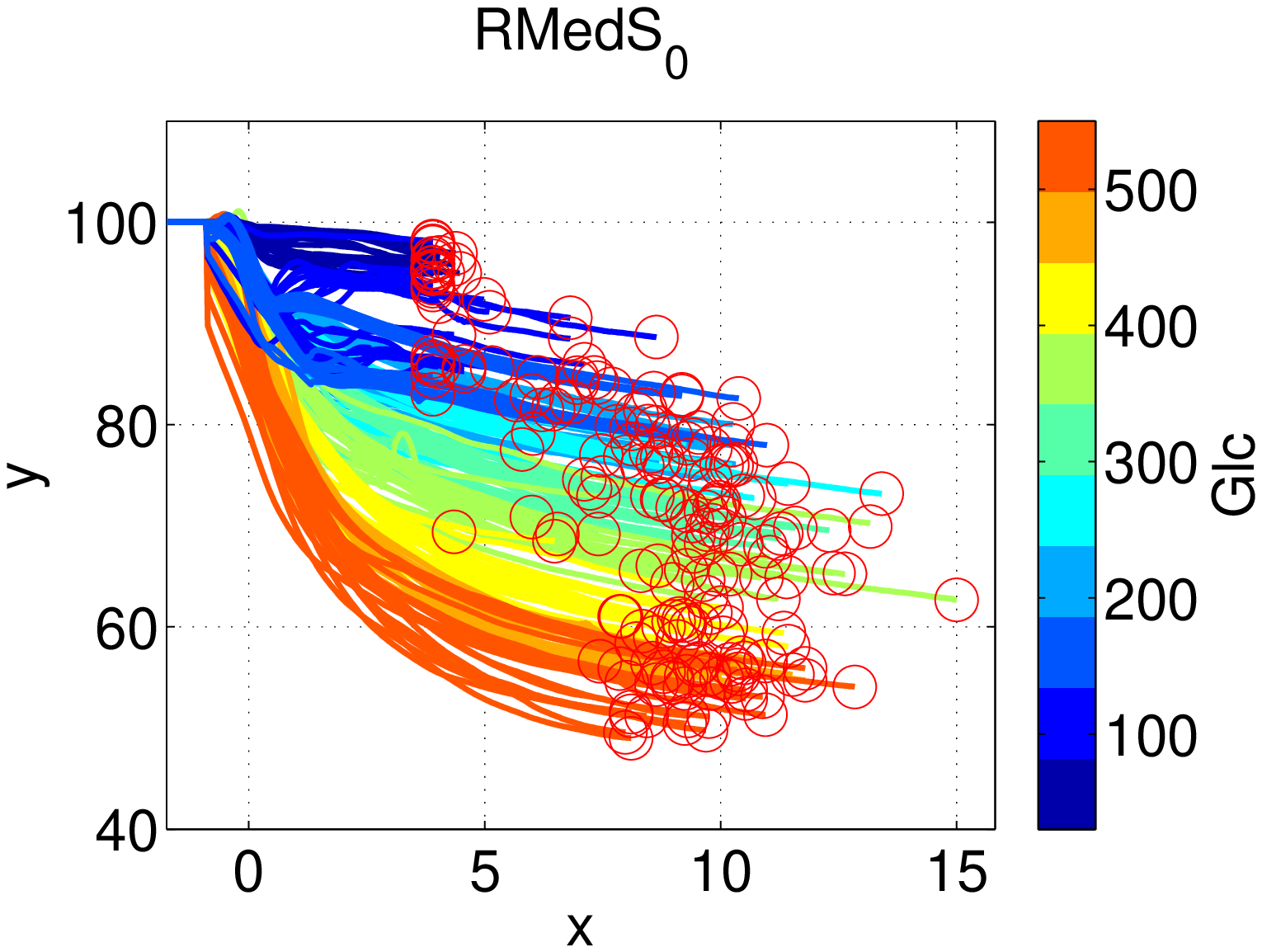}}
\vspace{-3pt}
\centerline{(c) Set E using MedS}\medskip
\end{minipage}
\hfill
\begin{minipage}[b]{.49\linewidth}
\centering
\psfrag{y}[rc]{\scriptsize $\hat{r}$ $(\%)$}
\psfrag{x}[bc]{\scriptsize Time [s]}
\psfrag{Glc}[c]{\scriptsize Glucose Level [mg/dl]}
\psfrag{title}[bc]{}
\centerline{\includegraphics[width=1\linewidth]{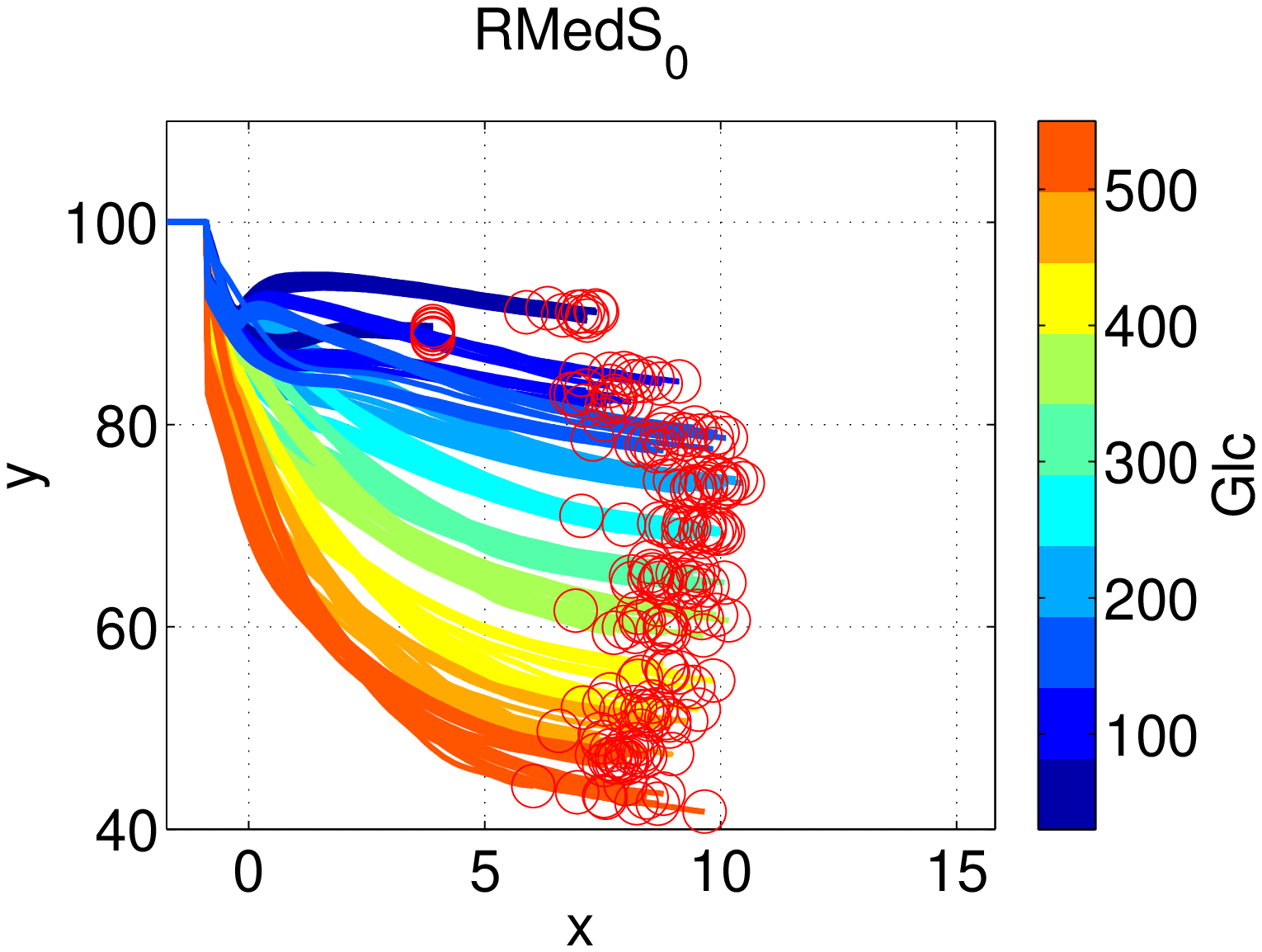}}
\vspace{-3pt}
\centerline{(d) Set F using R-MedS}\medskip
\end{minipage}
\hfill
\caption{Kinetic curves for different Sets using the different variants of MS and MedS. The red circles signify convergence points $\hat{r}_C^{(\text{Stand. Conv})}$. }
\label{fig:kinetic_results}
\end{figure}

In Fig.~\ref{fig:kinetic_results}, we present a selection of the results of the kinetic curves, i.e., the progression of the relative remission over time for the different data sets, using MS (mean-shift), RSS-MS (robust sparse MS), MedS (medoid-shift), and R-MedS (robust MedS). 
For the time being, we will present the results using the standard convergence criterion, i.e., when the slope reaches a threshold of $T_{\text{slope}}$ for a series of consecutive frames. 
Overall, we establish the expected progression of the kinetic curve with its three typical stages as in Fig.~\ref{fig:kinetik}.
We see that measurements of the same underlying glucose concentration are, mostly,  bundled together. Convergence is reached faster for low glucose concentrations than for high glucose concentrations. 
Moreover, our assumption that the decay rate $\tau$ of the reaction is correlated with the underlying glucose concentration is confirmed. The higher the underlying glucose concentration, the steeper the decay. 
\\ We observe a significant decrease in performance when comparing Set E with Set F, which were taken using the exact same setup and blood samples, however using different volumes of the blood samples.
\begin{table*}[ht!]
\begin{center}\caption{$\text{CV}_{\hat{r}}$ values for the different data sets using the standard convergence criterion.}.
\small\begin{tabular}{p{10pt} p{35pt}p{35pt}p{35pt}p{35pt}p{35pt} p{35pt}p{35pt}p{35pt}p{35pt}}
   \toprule
Set  &{MS}&{R-MS}&{MedS}&{R-MedS}& &{SS-MS}&{RSS-MS}&{SS-MedS}&RSS-MedS\\\cmidrule{1-5}\cmidrule{7-10}
A &1.59 & 1.29 & 1.17 & 1.22 && 1.87 & 1.81 & 1.52 & 1.52  \\
B & 1.95& 1.83 &1.82 &1.71 &&2.4 & 2.1 & 1.95 &1.95 \\
C & 1.15 &1.51 &0.90 &0.99&&1.50 & 1.41 & 1.39  &1.39\\
D & 1.87 & 1.7  & 1.58 & 1.39& & 1.87 & 2.90 &1.10 &1.09  \\
E &3.98 & 4.17& 3.39 & 3.25 & &5.9 & 5.6 & 4.3 & 4.2 \\
F & 1.48 & 1.37 & 1.10 & 1.04 & &2.44&2.25 &1.99  &1.78 \\
    \bottomrule
\end{tabular}
\label{tab:CVR_results}
\end{center}
\end{table*}
\begin{figure}[hbt]
\psfrag{70}[tc]{\scriptsize 70} \psfrag{80}[tc]{\scriptsize 80} \psfrag{90}[tc]{\scriptsize 90} \psfrag{100}[tc]{\scriptsize 100} \psfrag{50}[tc]{\scriptsize 50} \psfrag{60}[tc]{\scriptsize 60}\psfrag{110}[tc]{\scriptsize 110}\psfrag{120}[tc]{\scriptsize 120}
\psfrag{30}[tc]{\scriptsize 30}\psfrag{40}[tc]{\scriptsize 40}\psfrag{500}[br]{}\psfrag{2000}[br]{}\psfrag{2500}[br]{}\psfrag{3000}[br]{}
\psfrag{0}[tc]{\scriptsize }\psfrag{1000}[br]{}\psfrag{1500}[br]{}\psfrag{3500}[br]{\scriptsize}\psfrag{y}[bb]{\scriptsize Occurrence}
\psfrag{x}[t]{\scriptsize Intensity Value $(\%)$}
\begin{minipage}[b]{0.49\linewidth}
\centering
\psfrag{MS}[bc]{\scriptsize MS kinetic Curves}
\centerline{\includegraphics[width=1\linewidth]{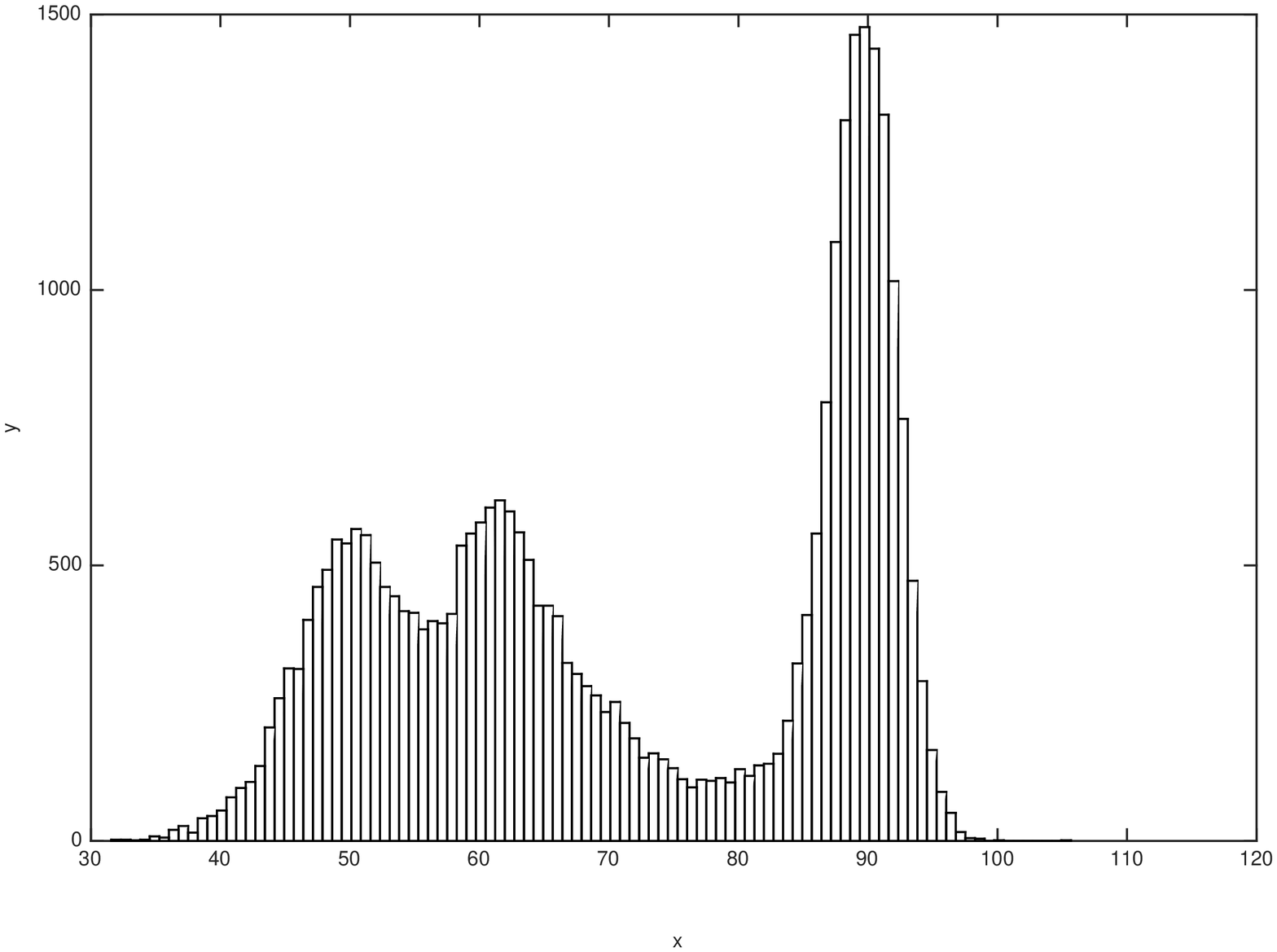}}
\vspace{7pt}
\centerline{(a)}\medskip
\end{minipage}
\begin{minipage}[b]{.49\linewidth}
\centering
\psfrag{MS}[bc]{\scriptsize State Estimation for MS}
\centerline{\includegraphics[width=1\linewidth]{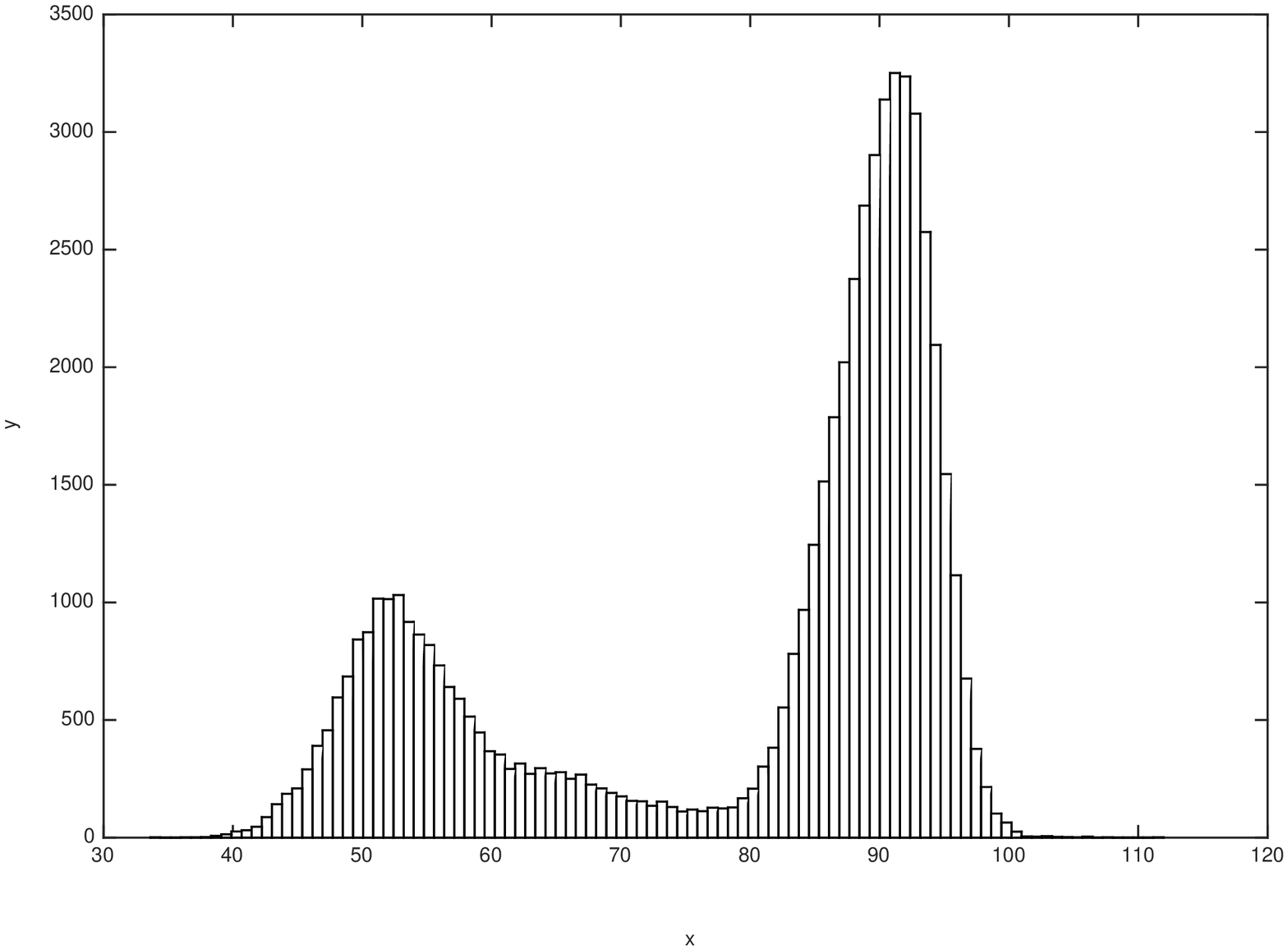}}
\vspace{7pt}
\centerline{(b)}\medskip
\end{minipage}
\hfill
\caption{Histograms of converged images of high glucose range measurements from (a) Set B, (b) Set C.}
\label{fig:hist_meanmedcomp}
\end{figure}
These results are underlined when analyzing the resulting $\text{CV}_{\hat{r}}$ values for these two sets in the left part of Table~\ref{tab:CVR_results}. This indicates that blood sample volumes in the range of $\SI{1}{nl} $ could be problematic.
The volume range $\SIrange{10}{100}{nl}$, in contrast, seems to perform well.\\Table~\ref{tab:CVR_results} shows that the medoid-shift versions are significantly better than the mean-shift versions for all sets.
Observing Set B, we notice that the difference in performance between the two methods is not as high as for other sets. The reason can be understood from Fig.~\ref{fig:hist_meanmedcomp}. The more distinct the modes are, the more separable the clusters, the better mean-shift performs. By choosing the medoid instead of the mean and converging to actual points in the data set, the medoid-shift doesn't show this bias in estimators.
\\ The results for the sparse scalable mean-shift are given in the right part of Table~\ref{tab:CVR_results}. We assert that the resulting $\text{CV}_{\hat{r}}$ values are slightly worse than their non-scalable counterparts. This degradation, however, is not too severe and the loss in accuracy can be accepted to ensure lower computation. Only for Set E do we notice a severe degradation in performance, when using the SS-MS. This can be traced back to the fact that due to the very small blood volume, the ROI is much smaller than for the other measurements. 
\subsubsection{Data-driven Choice of $N_\nu$ for the Sparse MS}
\begin{figure}[hbt]
\psfrag{30}[tc]{\scriptsize 30} \psfrag{60}[tc]{\scriptsize 60}\psfrag{75}[tc]{\scriptsize 75} \psfrag{90}[tc]{\scriptsize 90} \psfrag{105}[tc]{\scriptsize 105} \psfrag{120}[tc]{\scriptsize 120} \psfrag{135}[tc]{\scriptsize 135}\psfrag{150}[tc]{\scriptsize 150}
\psfrag{180}[tc]{\scriptsize 180}\psfrag{220}[tc]{\scriptsize 220}\psfrag{260}[tc]{\scriptsize 260}\psfrag{300}[tc]{\scriptsize 300}\psfrag{350}[tc]{\scriptsize 350}\psfrag{400}[tc]{\scriptsize 400}\psfrag{450}[tc]{\scriptsize 450}\psfrag{550}[tc]{\scriptsize 550}
\psfrag{200}[tc]{\scriptsize 200}\psfrag{250}[tc]{\scriptsize 250}
\psfrag{100}[cc]{\scriptsize 100}
\psfrag{x}[t]{\scriptsize Glucose Concentration [mg/dl]}\psfrag{y}[rb]{\scriptsize $N_\nu$}
\centerline{\includegraphics[width=1\linewidth]{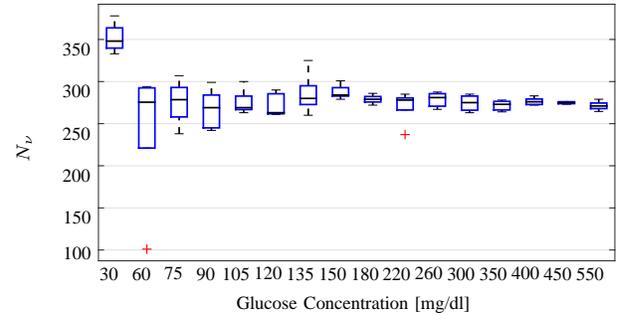}}
\caption{The choice of $N_\nu$ for different glucose concentrations for Set C.}
\label{fig:kanalyse}
\end{figure}

As portrayed in Section~\ref{sec:SS-MS}, we propose a data-driven approach to select $N_\nu$ uniquely for every frame. Figure.~\ref{fig:kanalyse} shows the choice of $N_\nu$ for Set C, with $N_\nu$ averaged over all frames for each measurement.
The tendency matches our expectations:
very low glucose concentrations are characterized by low contrast and therefore higher coherence. The choice of $N_\nu$ is higher than for high glucose concentrations, i.e., more data points are needed
to reliably represent the image. As the concentration increases
and the images become more distinct in the different regions, the variance in the choice of $N_\nu$ for the same glucose concentrations decreases. Taking the worst-case scenario would have resulted in $N\approx 370$ data points, which is much higher than the data-driven choice of $N$.
\\ Let us now analyze the quality of the SS-MS when using a  data-driven choice of $N_\nu$ versus a fixed $N_\nu = 0.37L$. The results are given for the RSS-MedS in Table.~\ref{tab:barplots_scalable}. Evidently, the accuracy of both approaches is comparable. This underlines that
a data-driven $N_\nu$ maintains the accuracy needed.
\begin{table}[htb]
\centering\caption{Comparison of the $\text{CV}_{\hat{r}}$ values for a data-driven choice of $N_\nu$ vs. a fixed $N_\nu = 0.37L$. }
\rowcolors{2}{gray!25}{white}
\small\begin{tabular}{ c c c c c c c }
\rowcolor{gray!50}
\toprule
Choice of $N_\nu$ & A & B & C & D & E & F\\\midrule
Data-driven & 1.03 & 1.29 & 0.81 & 1.90 & 4.60 & 1.59\\\hline 
Fixed & 0.95 & 1.38 & 1.27 & 1.97 &4.35 &1.64\\ \bottomrule 
\end{tabular}\vspace{10pt}
\label{tab:barplots_scalable}\end{table}
\subsubsection{State Estimation using the EKF}
The kinetic curves in Fig.~\ref{fig:kinetic_results} show that convergence for same-glucose-level measurements is not always reached at the same time which results in strongly varying remission estimates, although the course of the kinetic curve is very similar. 
\begin{figure}[hbt]
\psfrag{70}[bl]{\scriptsize 70} \psfrag{80}[bl]{\scriptsize 80} \psfrag{90}[bl]{\scriptsize 90} \psfrag{100}[bl]{\scriptsize 100} \psfrag{50}[bl]{\scriptsize 50} \psfrag{60}[bl]{\scriptsize 60}
\psfrag{200}[bl]{\scriptsize 200}\psfrag{300}[bl]{\scriptsize 300}\psfrag{400}[bl]{\scriptsize 400}\psfrag{500}[bl]{\scriptsize 500}
\psfrag{140}[bl]{\scriptsize 140}\psfrag{130}[bl]{\scriptsize 130}\psfrag{120}[bl]{\scriptsize 120}\psfrag{110}[bl]{\scriptsize 110}
\psfrag{0}[bl]{\scriptsize 0}\psfrag{5}[bl]{\scriptsize 5} \psfrag{10}[bl]{\scriptsize 10} \psfrag{15}[bl]{\scriptsize 15} 
\begin{minipage}[b]{0.48\linewidth}
\centering
\psfrag{y}[bc]{\scriptsize $\hat{r}_C [\%]$}
\psfrag{x}[cc]{\scriptsize Time $[s]$}
\psfrag{title}[bc]{}
\psfrag{MS}[bc]{\scriptsize MS kinetic curves}
\psfrag{Glc}[bc]{\scriptsize Glucose Level [mg/dl]}
\centerline{\includegraphics[width=.99\linewidth]{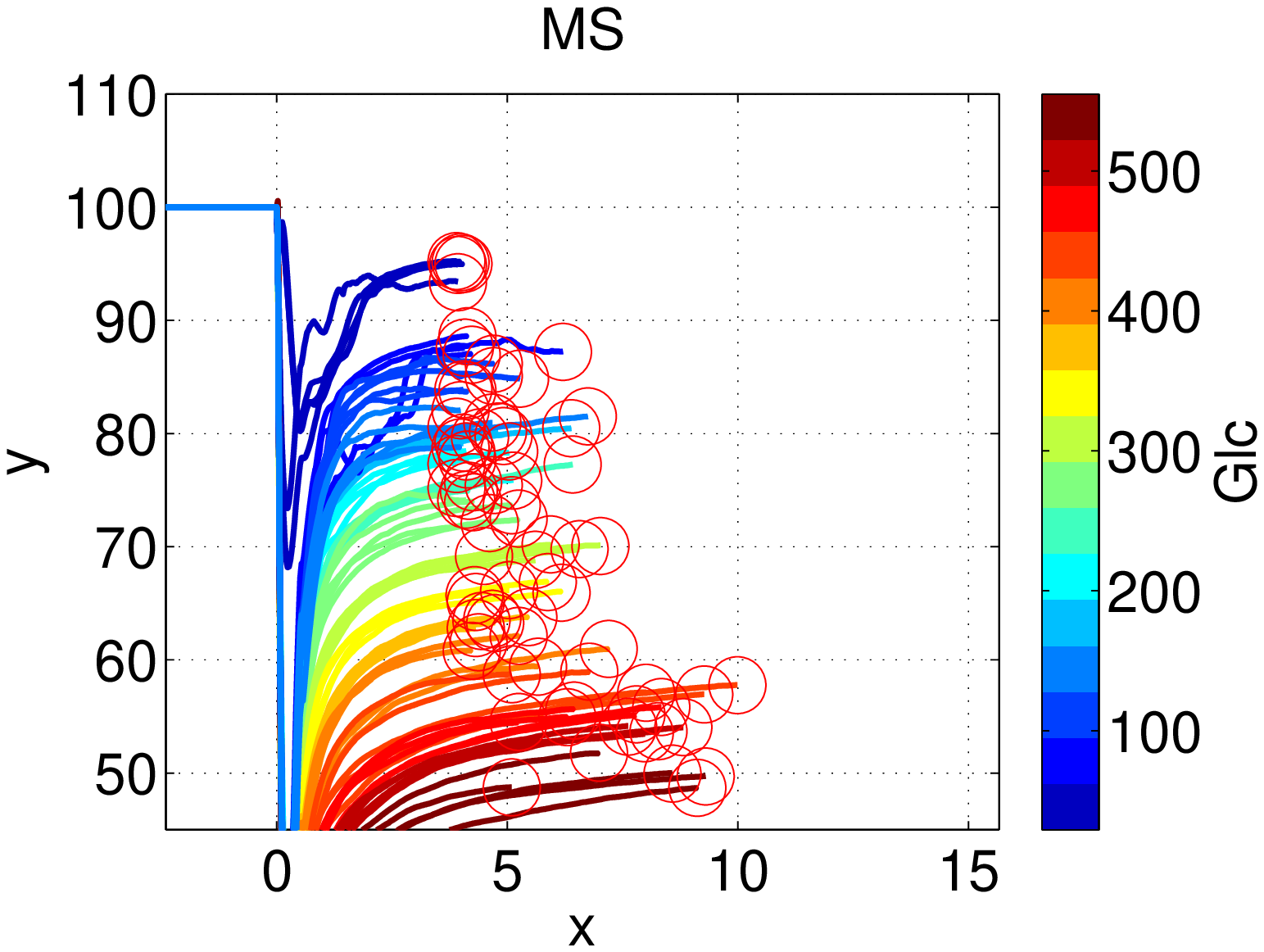}}
\vspace{-3pt}
\centerline{(a)}\medskip
\end{minipage}
\begin{minipage}[b]{.48\linewidth}
\centering
\psfrag{Kinetic}[lc]{\scriptsize Kinetic curve}
\psfrag{RCkin}[lc]{\scriptsize $\hat{r}_C^{(\text{Stand. Conv})}$}
\psfrag{RCkalm}[lc]{\scriptsize $\hat{r}_C^{(\text{EKF Conv})}$}
\psfrag{x}[cc]{\scriptsize Time $[s]$}
\psfrag{title}[bc]{}
\psfrag{y}[bc]{\scriptsize $\hat{r} [\%]$}
\psfrag{Glc}[bc]{\scriptsize Glucose Level [mg/dl]}
\psfrag{MS}[bc]{\scriptsize State Estimation for MS}
\centerline{\includegraphics[width=.99\linewidth]{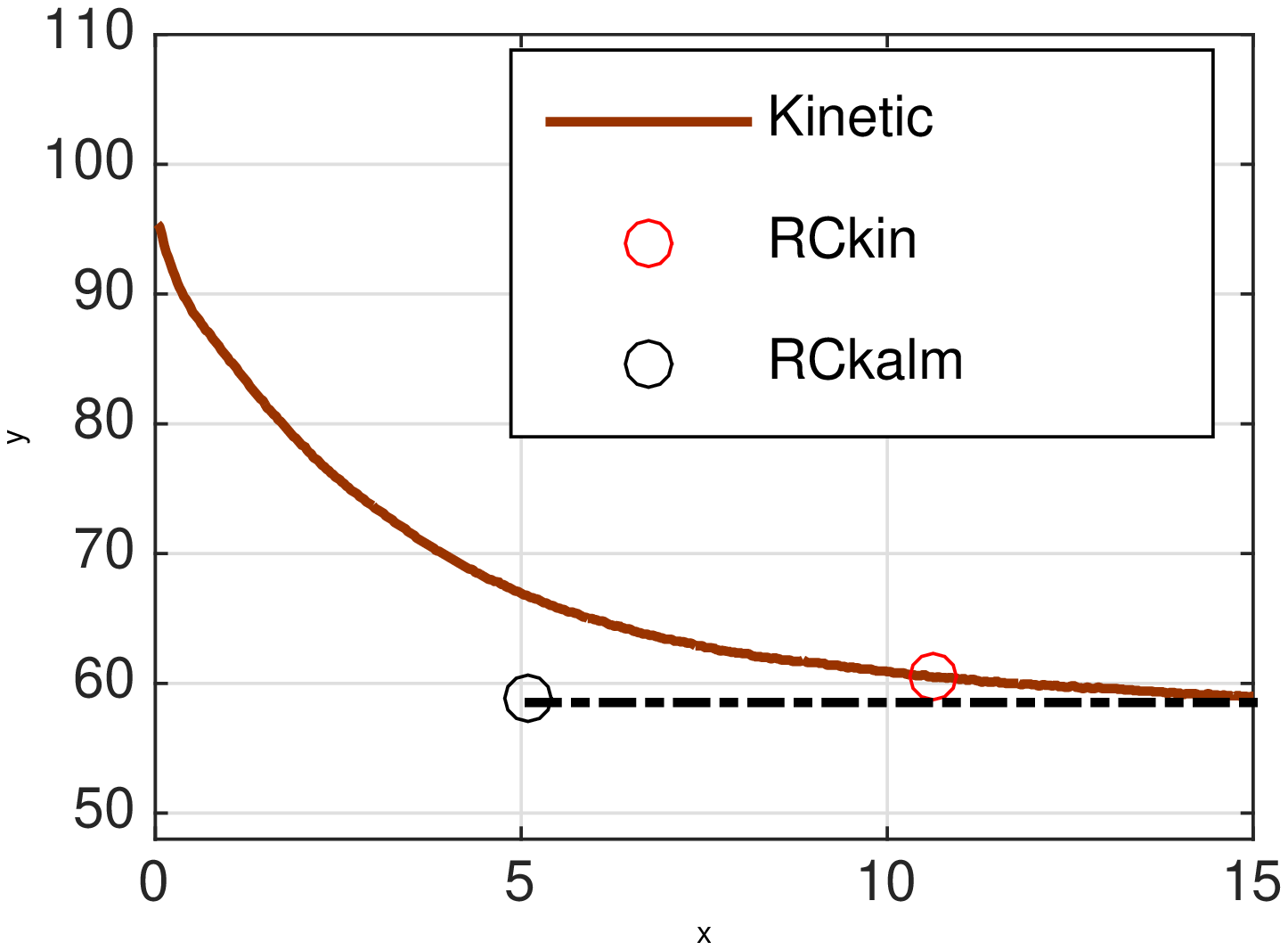}}
\centerline{(b)}\medskip
\end{minipage}
\hfill
\caption{(a) State estimation of $r_C$ using the EKF. The red circles ($\circ$) indicate the estimated convergence values. (b) A comparison of the convergence estimates using the standard convergence criterion and the EKF.}
\label{fig:Kalmanresults}
\end{figure}
Let us take the $\SI{550}{mg/dl}$ measurements in Fig.~\ref{fig:kinetic_results} (b) as an example. Convergence is found at times between $\hat{t}_C \approx \SI{4}{s}$ and $\hat{t}_C \approx \SI{9}{s}$, leading to remission estimates from $\hat{r}_C \approx 50 \%$ to $\hat{r}_C \approx 59 \%$.
Figure~\ref{fig:Kalmanresults} (a) shows the EKF state estimates $\hat{r}_C^{(\text{EKF Conv})}$ for this example. Evidently, the EKF converges to reliable state estimates quickly and therefore, the resulting 
$\hat{r}_C$ estimates not only lie in the same range, but match the final convergence value of the kinetic behavior more accurately. This can be seen clearly in Fig.~\ref{fig:Kalmanresults} (b). 
The convergence value reached by the standard method $\hat{r}_C^{(\text{Stand. Conv})}$ occurs after around $\SI{11}{s}$ and does not match the actual saturation value. The convergence value reached by EKF $\hat{r}_C^{(\text{EKF Conv})}$, however, occurs after $\SI{5}{s}$ and is much closer to the actual saturation value.
\begin{table}[htb]
\centering\caption{Time gain and reduced error obtained using EKF.}
\small
\begin{tabular}{ccc}
\toprule
   &Time Gain& Error Benefit\\ \cmidrule{2-3}
Low Glucose Range ($g\leq \SI{75}{mg/dl}$) & \SI{1.41}{s}&2.23 \% \\
High Glucose Range ($g>\SI{75}{mg/dl}$) & \SI{0.97}{s}& 2.1 \%\\\bottomrule
\end{tabular}
\label{tab:EKF_benefits}\end{table}
\begin{figure}[hbt]
\centering
\psfrag{conv2}[cl]{\scriptsize EKF}\psfrag{conv1}[cl]{\scriptsize Stand.}
\psfrag{A}[tc]{\scriptsize A}\psfrag{B}[tc]{\scriptsize B}\psfrag{C}[tc]{\scriptsize C}
\psfrag{D}[tc]{\scriptsize D}\psfrag{E}[tc]{\scriptsize E}\psfrag{F}[tc]{\scriptsize F} \psfrag{Sets}[tc]{\scriptsize Sets}
\psfrag{CVR}[br]{\scriptsize $\text{CV}_{\hat{r}}$}
\psfrag{1.5}[br]{\scriptsize $1.5$}\psfrag{2.5}[br]{\scriptsize$2.5$}\psfrag{3.5}[br]{\scriptsize$3.5$}\psfrag{0.5}[br]{\scriptsize$0.5$}\psfrag{0}[br]{\scriptsize$0$}\psfrag{1}[br]{\scriptsize$1$}\psfrag{2}[br]{\scriptsize$2$}\psfrag{3}[br]{\scriptsize$3$}
\psfrag{conv2low}[cl]{\scriptsize EKF - Low glc}\psfrag{conv1low}[cl]{\scriptsize Stand. - Low glc}
\psfrag{conv2high}[cl]{\scriptsize EKF - High glc}\psfrag{conv1high}[cl]{\scriptsize Stand. - High glc}
\psfrag{A}[tc]{\scriptsize A}\psfrag{B}[tc]{\scriptsize B}\psfrag{C}[tc]{\scriptsize C}
\psfrag{D}[tc]{\scriptsize D}\psfrag{E}[tc]{\scriptsize E}\psfrag{F}[tc]{\scriptsize F} \psfrag{Sets}[tc]{\scriptsize Sets}
\psfrag{gmse}[br]{\scriptsize gMAD}
\psfrag{5}[br]{\scriptsize$5$}\psfrag{15}[br]{\scriptsize$15$}\psfrag{10}[br]{\scriptsize $10$}\psfrag{25}[br]{\scriptsize$25$}\psfrag{0}[br]{\scriptsize$0$}\psfrag{20}[br]{\scriptsize$20$}\psfrag{35}[br]{\scriptsize$35$}\psfrag{30}[br]{\scriptsize$30$}\psfrag{3}[br]{\scriptsize$3$}
\begin{minipage}[b]{0.49\linewidth}
\centerline{\includegraphics[width=.8 \linewidth]{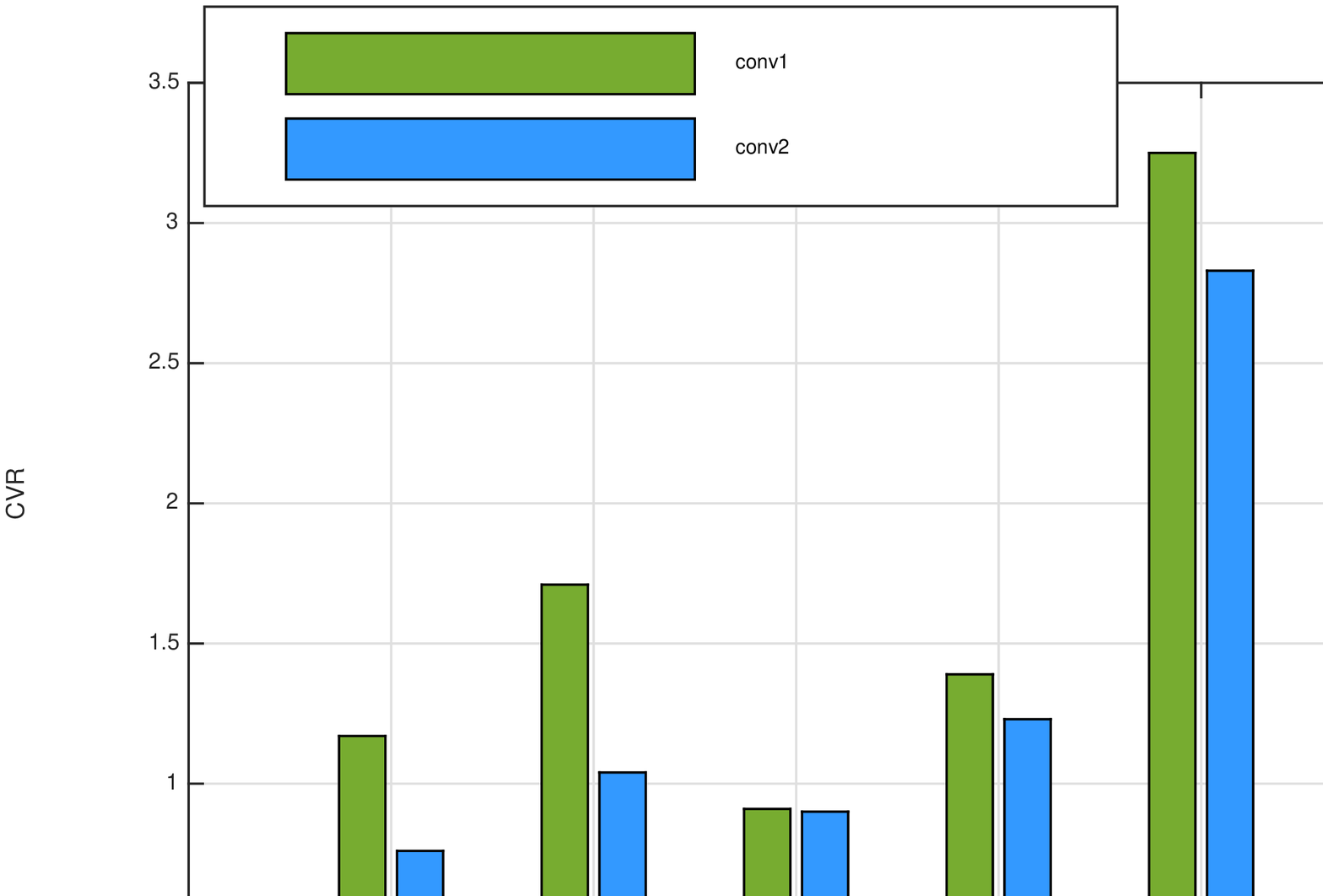}}
\vspace{15pt}
\centerline{(a)}\medskip
\end{minipage}
\hfill
\begin{minipage}[b]{0.49\linewidth}
\centerline{\includegraphics[width=.8\linewidth]{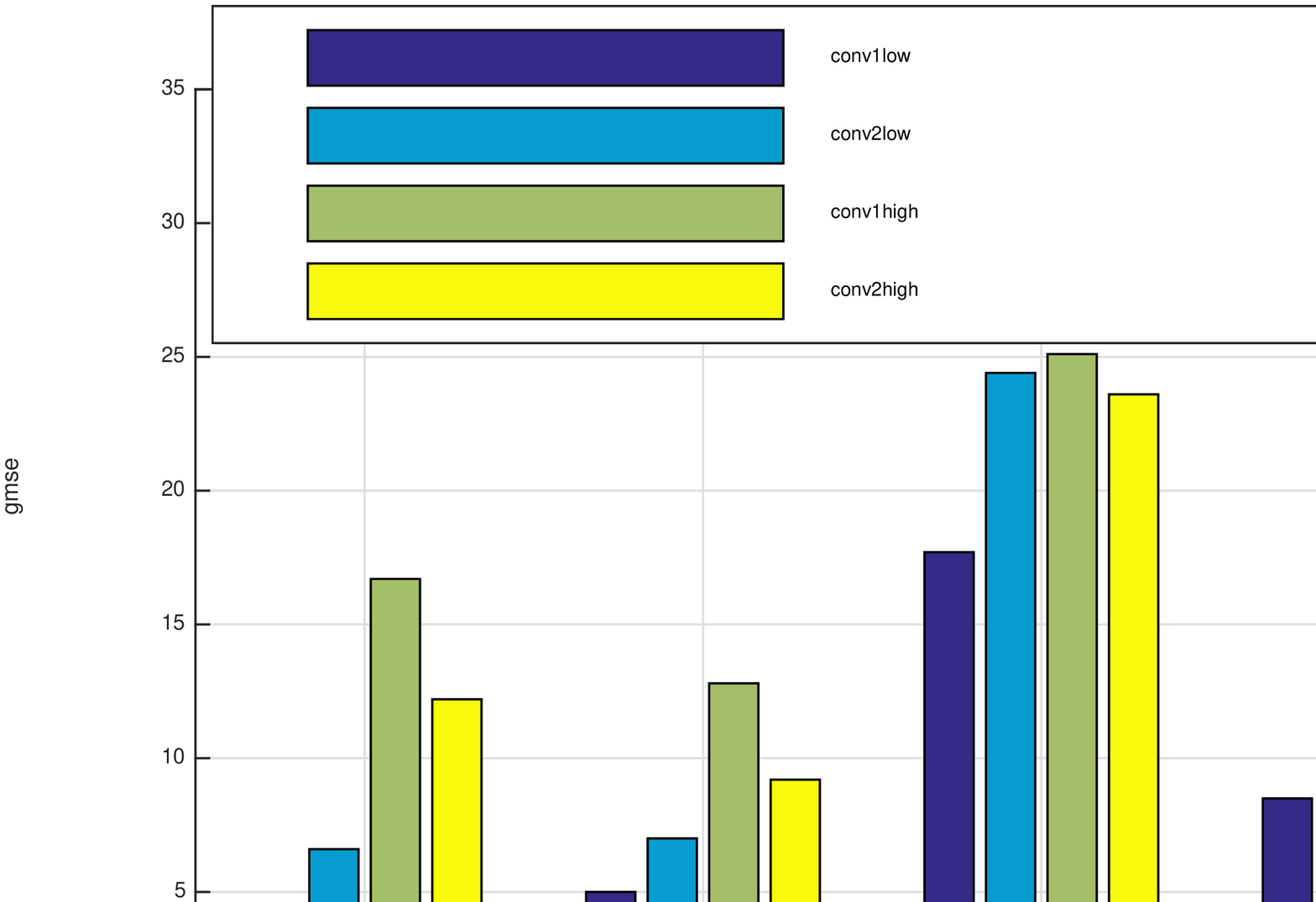}}
\vspace{15pt}
\centerline{(b)}\medskip
\end{minipage}
\hfill
\caption{(a) Comparison of the $\text{CV}_{\hat{r}}$ values for the standard convergence criterion to EKF. (b) Comparison of the gMAD values for the standard convergence criterion to EKF.}
\label{fig:EKF_barplot}
\end{figure}
\\ Table \ref{tab:EKF_benefits} shows the time gain obtained using the EKF. These results 
are the average of all time gain values obtained for all sets over all measurements. Comparing the remission estimates 
$\hat{r}_C^{(\text{Stand. Conv})}$ and $\hat{r}_C^{(\text{EKF Conv})}$ to the saturation values of the chemical reaction, we assert that the EKF produces more accurate results, with a gain of more than $2\%$ in remission. 
\\Figure~\ref{fig:EKF_barplot} presents a comparison of $\text{CV}_{\hat{r}}$ values and gMAD values for the standard convergence criterion and the EKF convergence method. 
This analysis considers for each set the method with the best results in Tables~\ref{tab:CVR_results} and \ref{tab:gmse_results_kinconv}, respectively,  and compares it to its EKF-method-equivalent. 
The EKF convergence always leads to improved $\text{CV}_{\hat{r}}$ values. Similarly, for the high glucose range the gMAD results show the same outcome. For low glucose ranges the standard convergence compares favourably. 
The reason for this is that the EKF method is quite sensitive to the model used. As can be seen in Fig.~\ref{fig:kinetic_results}, the kinetic curves of low glucose measurements exhibit a dip after the drop followed by a steady rise. 
This is not embodied in the model in Eq.~(\ref{eq:model2}).
\subsubsection{Mapping to the Underlying Glucose Concentrations}
Finally, we present the results of the glucose estimates $\hat{g}$ obtained by mapping the remission estimates using $f_\text{Calib}$. 
\begin{figure}[h!]
\psfrag{x}[bl]{\scriptsize $g$ [mg/dl]}\psfrag{y}[bl]{\scriptsize $\hat{g}$ [mg/dl]}\psfrag{Clarke's Error Grid Analysis}[bl]{}
\psfrag{E}[bl]{\scriptsize $E$}\psfrag{A}[bl]{\scriptsize $A$}\psfrag{B}[bl]{\scriptsize $B$}\psfrag{C}[bl]{\scriptsize $C$}\psfrag{D}[bl]{\scriptsize $D$}
\psfrag{600}[bl]{\scriptsize $600$}\psfrag{500}[bl]{\scriptsize $500$}\psfrag{400}[bl]{\scriptsize $400$}\psfrag{300}[bl]{\scriptsize $300$}\psfrag{200}[bl]{\scriptsize $200$}\psfrag{100}[bl]{\scriptsize$100$}
\psfrag{0}[bl]{\scriptsize $0$}\psfrag{x}[tc]{\scriptsize $g$ [mg/dl]}\psfrag{title}[bc]{\scriptsize MS}\psfrag{y}[Bc]{\scriptsize $\hat{g}$ [mg/dl]}
\psfrag{70}[bl]{\scriptsize 70} \psfrag{80}[bl]{\scriptsize 80} \psfrag{90}[bl]{\scriptsize 90} \psfrag{50}[bl]{\scriptsize 50} \psfrag{60}[bl]{\scriptsize 60}
\psfrag{140}[bl]{\scriptsize 140}\psfrag{130}[bl]{\scriptsize 130}\psfrag{120}[bl]{\scriptsize 120}\psfrag{110}[bl]{\scriptsize 110}
\begin{minipage}[b]{.49\linewidth}
\centering
\psfrag{title}[bc]{}
\centerline{\includegraphics[width=1\linewidth]{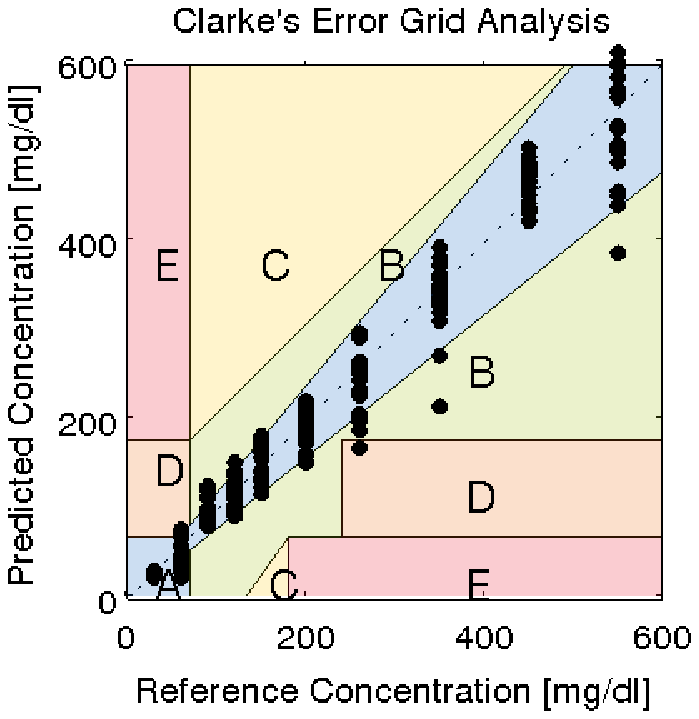}}
\vspace{-3pt}
\centerline{(a)}\medskip
\end{minipage}
\hfill
\begin{minipage}[b]{.49\linewidth}
\centering
\psfrag{title}[bc]{}
\centerline{\includegraphics[width=1\linewidth]{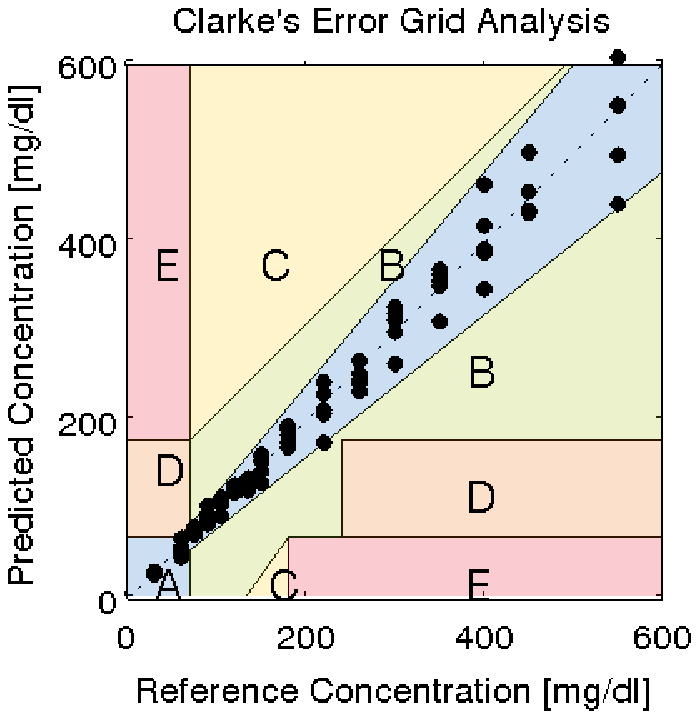}}
\vspace{-3pt}
\centerline{(b)}\medskip
\end{minipage}
\hfill
\caption{Clarke's Error Grid Analysis for Set E for (a) the standard and (b) the EKF convergence criteria. }
\label{fig:CEG_results}
\end{figure}
The Clarke Error Grid plot of Set E in Fig.~\ref{fig:CEG_results} (a) shows that 93 $\%$ of the points lie in region A, 6 $\%$ lie in region B and $1\%$ lie in region D using the standard convergence criterion. Thereby, the results do not conform with the requirements.
As indicated by the results in Fig.~\ref{fig:CEG_results} (b), using EKF state estimation to predict the convergence values leads to a significant improvement of the results. Now 100$\%$ of the estimates of Set E lie in the A-region. 
\begin{table}[ht]
\begin{center}
\small
\begin{tabular}{p{8pt} p{17pt}p{17pt}p{18pt}p{20pt} p{17pt}p{17pt}p{18pt}p{20pt}}
    \toprule
  &\multicolumn{4}{c}{Low Glucose ($\leq\SI{75}{mg/dl}$) }&\multicolumn{4}{c}{High Glucose ($>\SI{75}{mg/dl}$) }\\\cmidrule(lr){2-5} \cmidrule(lr){6-9}
  Set  &{MS}&{R-MS}&{MedS}&{R-MedS}&{MS}&{R-MS}&{MedS}&{R-MedS}\\\cmidrule(r){1-5}\cmidrule(lr){6-9}   
B & 14.4 & 13.0 & \bf{6.6} & \bf{6.6} &13.4 &\bf{11.8} &17.0 &17.2 \\
C & 6.9 &6.9 &4.6& \bf{4.0}& 12.8 & 14.4 & \bf{7.8} & 8.9  \\
E & 22.6 & 29.4 & 24.4 &\bf{21.0}&30.8 &37.4 &25.5 &\bf{23.6}  \\
F & 9.3 & 9.4 &  8.5 &\bf{8.1}& 17.5 &17.0 & \bf{16.6} &18.4  \\
    \bottomrule
\end{tabular}\caption{gMAD values using the standard convergence criterion.}.
\label{tab:gmse_results_kinconv}
\end{center}
\end{table}
Table~\ref{tab:gmse_results_kinconv} presents the mean gMAD values for $\hat{g}$ of each set separated according to the glucose range. This analysis is omitted for Sets A and D, as they do not contain enough different 
glucose concentrations to construct the mapping function. The results for most sets lie below the ISO requirements, i.e., low glucose ranges show errors smaller than gMAD~$= \pm\SI{15}{mg/dl}$ and high glucose ranges show an error larger than 20$\%$. However, we note that in Set E the limit is often exceeded, confirming again the fact that blood sample volumes around $\SI{1}{nl}$ are problematic.
\\The previously made observation concerning medoid-shift outperforming mean-shift is confirmed again here. The robust version seems to improve the performance in high glucose ranges more than in low glucose ranges. 
\section{Conclusion}
\label{sec:conclusion}
Regular self-control using hand-held glucometers is an indispensable part of diabetes care and therefore, glucometers should maintain high accuracy while improving usability.
We have developed a full framework to measure the blood glucose concentration from glucose images using blood sample volumes in the nl-range, which is much smaller than the state-of-the-art, while complying with the most recent ISO standards for accuracy\cite{ISO2013}.
Using the mean-shift principle and its variations, the robust and scalable sparse mean-shift, the intensity level of the region of interest is estimated. 
We have shown that the scalable version of the mean-shift with an individual selection of the number of data points gives good results w.r.t. accuracy, while decreasing the computation time. 
These variations are extended to the medoid-shift, which outperforms the mean-shift in our experiments. Furthermore, the convergence of all mean-shift and medoid-shift variations is proven.
We assert that the mean-shift and its variations are suitable for segmentation applications, where the number and size of the regions is unknown.
The extended Kalman filter and a model for the chemical reaction are employed to enhance the accuracy of the estimate and reduce the measurement time by around $20\%$. A linear relation has been found between the convergence of the glucose
reaction and its decay rate.
We establish that inaccuracies in the derived model can lead to degraded performance and, therefore, the model has to be calibrated uniquely for each specific setup used. 
As future work, we aim to enhance this model, taking into account not only the underlying 
glucose concentration but also incorporating affecting parameters such as temperature, humidity, and the haematocrit level \cite{Shin2003} and viscosity of the sample. Additionally, we see the need to further validate our framework with clinical tests. Finally, we establish that very low blood sample volumes of $\SI{1}{nl}$ seem to be at the limit of what is acceptable in accuracy.
\vspace{-9pt}
\appendices %
\section{Proof of convergence of the mean-shift algorithm with weights $w_l$}%
\label{sec:app1}
\begin{proof}
Since $L$ is finite, the sequence $\{\func{\hat{f}_K}{\x^{(j)}}\}$, ${ j=1,2,\ldots}$ is bounded. We will show that for $\x^{(j)}\neq\x^{(j+1)}$: $\func{\hat{f}_K}{\x^{(j)}}<\func{\hat{f}_K}{\x^{(j+1)}}$. Without loss of generality we assume $\x^{(j)}=0$, $k$ being the profile of the kernel $\func{K}{\cdot}$ \cite{Comaniciu2002} 
\vspace{-10pt}
\begin{align}\label{eq:proof1}
\func{\hat{f}_K}{\x^{(j+1)}}&-\func{\hat{f}_K}{\x^{(j)}} = \frac{1}{\h} \sum_{l=1}^{L} w_l\\& \times\left[
k\left(\frac{\x^{(j+1)}-x_l}{h}\right)- \func{k}{\frac{x_l}{h}}\right]\nonumber
\end{align}
For convex profiles and $x_2 \neq x_1, x_1,x_2 \in [0,\infty)$ it follows $\func{k}{x_2}\geq \func{k}{x_1} + \func{k'}{x_1}(x_2-x_1)$, such that
\begin{align}
\func{\hat{f}_K}{\x^{(j+1)}}&-\func{\hat{f}_K}{\x^{(j)}} \geq -\frac{1}{\h} \sum_{l=1}^{L} w_l \cdot \func{k'}{\frac{x_l}{h}}\cdot \\\nonumber
&\times\left[
\left|\left|\x_l\right|\right|^2-\left|\left|\x^{(j+1)}-\x_l\right|\right|^2\right]\\\nonumber
& = - 2\frac{1}{\h} \cdot \x^{T(j+1)}\sum_{l=1}^{L} w_l \x_l \cdot \func{k'}{\frac{x_l}{h}} \\\nonumber
& + \frac{1}{\h}\sum_{l=1}^{L} w_l \left|\left|\x^{(j+1)}\right|\right|^2 \cdot \func{k'}{\frac{x_l}{h}}\\\nonumber
& = -\sum_{l=1}^{L} w_l \left|\left|\x^{(j+1)}\right|\right|^2 \cdot \func{k'}{\frac{x_l}{h}}
\end{align}
Since $\func{k}{x}$ is monotonically decreasing, $-\func{k'}{x} \geq 0$ for $x \in [0,\infty)$. $\sum_{l=1}^{L}-\func{k'}{\frac{x_l}{h}}$ is strictly positive, as are the weights $w_l$. Hence $\func{\hat{f}_K}{\x^{(j+1)}}-\func{\hat{f}_K}{\x^{(j)}}>0$ and the sequence $\{\func{\hat{f}_K}{\x^{(j)}}\}$ is convergent.
\\ Without assuming $\x^{(j)}=0$ and reformulating (\ref{eq:proof1}) we get
\begin{align}
\func{\hat{f}_K}{\x^{(j+1)}}&-\func{\hat{f}_K}{\x^{(j)}} \geq \\
& = -\frac{1}{\h} \sum_{l=1}^{L} w_l \left|\left|\x^{(j+1)}-\x^{(j)}\right|\right|^2 \cdot \func{k'}{\frac{\x^{(j)}-x_l}{h}}\nonumber.
\end{align}
Since $\func{\hat{f}_K}{\x^{(j+1)}}-\func{\hat{f}_K}{\x^{(j)}} $ converges to zero, then $\left|\left|\x^{(j+1)}-\x^{(j)}\right|\right|^2$ also converges to zero and $\x^{(j)}$,${ j = 1,2,\ldots}$ is a Cauchy sequence.
\end{proof} 
\section{Extension of the proof of convergence for the medoid-shift with weights $w_l$}\label{sec:app_med}
\begin{proof}
The choice of successive points in the medoid-shift algorithm, as given in (\ref{eq:msvec2_med}), is carried out according to 
\begin{align}
& \sum_{l=1}^{L} \frac{1}{h}\Big|\Big|{\x^{(j)}-x_l}\Big|\Big|^2w_l \func{k'}{\frac{\x^{(j)}-x_l}{h}}\\\nonumber
 & >  \sum_{l=1}^L \frac{1}{\h} \Big|\Big|{\x^{(j+1)}-x_l}\Big|\Big|^2w_l\func{k'}{\frac{\x^{(j)}-x_l}{h}}, 
\end{align}
as equality of these two terms indicates convergence. This can be reformulated as
\begin{align}
&\sum_{l=1}^L \frac{1}{\h}w_l\func{k'}{\frac{\x^{(j)}-x_l}{h}} \\\nonumber
& \times\left[ \Big|\Big|{\x^{(j)}-x_l}\Big|\Big|^2- \Big|\Big|{\x^{(j+1)}-x_l}\Big|\Big|^2 \right]\\\nonumber
& = -\sum_{l=1}^L  \frac{1}{\h}w_l\Big|\Big|{\x^{(j+1)}-\x^{(j)}}\Big|\Big|^2\func{k'}{\frac{\x^{(j)}-x_l}{h}}  > 0,
\end{align}
\end{proof} 
proving that $\func{\hat{f}_K}{\x^{(j+1)}}>\func{\hat{f}_K}{\x^{(j)}} $, hence that the sequence $\{\func{\hat{f}_K}{\x^{(j)}}\}, j=1,2,\cdots$ is strictly positive and thus $\x^{(j)}\neq\x^{(j+c)}$, for all $c>0$. This proves that their are no cycles and medoid-shift will converge.
\section*{Acknowledgement}
The authors would like to thank Dipl.-Phys. B. Limburg from Roche Diagnostics GmbH, Mannheim, for his support and for providing the real glucose data.
\bibliographystyle{IEEEbib}

\vspace{-20mm}
\begin{IEEEbiography}[{\includegraphics[width=1in,height=1.25in,clip,keepaspectratio]{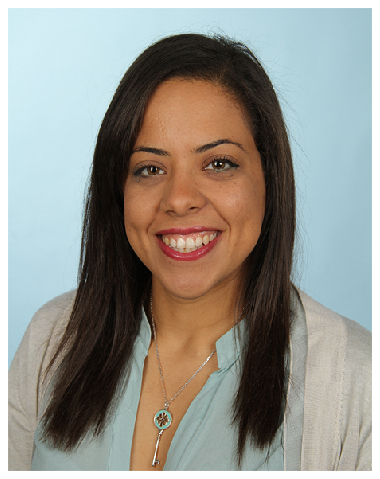}}]{Nevine Demitri}
 (S'11) was awarded a DAAD scholarship to pursue her B.Sc. and M.Sc. degrees in Technische Universit\"at Darmstadt, Germany. There, she received the B.Sc. in information and communication technology and M.Sc. in information technology and electrical engineering
 in 2008 and 2011, respectively. She is currently working towards the Ph.D. degree in the Institute of Telecommunications, Signal Processing Group, Technische Universit\"at Darmstadt. Her current research interests include biomedical signal processing, image and video processing,
 as well as machine learning.
 \end{IEEEbiography}
 \vspace{-20mm}
 \begin{IEEEbiography}[{\includegraphics[width=1in,height=1.25in,clip,keepaspectratio]{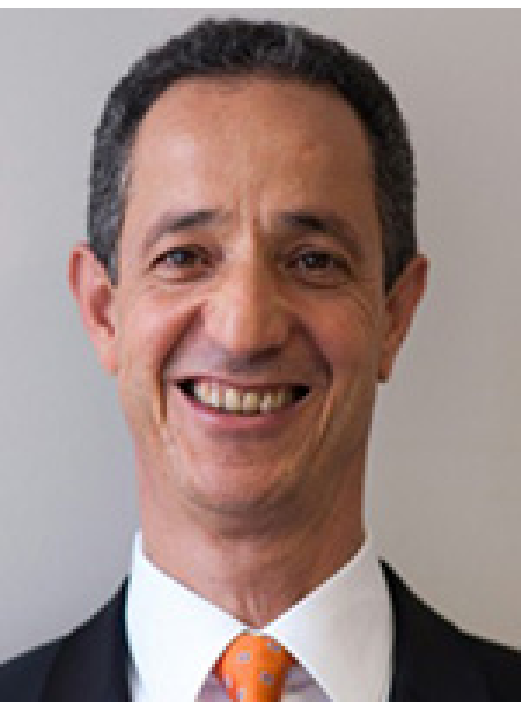}}]{Abdelhak M. Zoubir}
is a Fellow of the IEEE and IEEE Distinguished Lecturer (Class 2010- 2011). He received his Dr.-Ing. from Ruhr- Universit\"at Bochum, Germany, in 1992. He was with Queensland University of Technology, Australia, from 1992-1998 where he was Associate Professor. In 1999, he joined Curtin University of Technology, Australia, as a Professor of Telecommunications and was Interim Head of the School of Electrical \& Computer Engineering from 2001 until 2003. In 2003, he moved to Technische Universit\"at Darmstadt, Germany, as Professor of Signal Processing and Head of the Signal Processing Group. His research interest lies in statistical methods for signal processing with emphasis on bootstrap techniques, robust detection and estimation and array processing applied to telecommunications, radar, sonar, automotive monitoring and safety, and biomedicine. He published over 300 journal and conference papers on these areas. Professor Zoubir acted as General or Technical Chair of numerous conferences and workshops. 
Recently, he was the Technical Co-Chair of ICASSP-14 held in May in Florence, Italy. Dr Zoubir has also held several positions in editorial boards; most notably, he was the Editor-In-Chief of the IEEE Signal Processing Magazine (2012-2014). He was elected as Chair (2010-2011), Vice-Chair (2008-2009) and Member (2002-2008) of the IEEE SPS Technical Committee Signal Processing Theory and Methods (SPTM), and a Member (2007-2012) of the IEEE SPS Technical Committee Sensor Array and Multi-channel Signal Processing (SAM). He currently serves on the Board of Governors of the IEEE SPS as an elected Member-at-Large (2015-2017), and has been a Member of the Board of Directors of the European Association of Signal Processing (EURASIP) since 2008.
 \end{IEEEbiography}
\end{document}